\newlength{\intwidth}

\documentclass{jfm}

\usepackage{amssymb}

\usepackage{array}
\usepackage{amsmath}
\usepackage{latexsym}
\usepackage{bezier}
\usepackage{psfrag,graphicx}
\usepackage{bm}
\usepackage{float}
\usepackage{multirow}
\usepackage{mathrsfs}
\usepackage{color}
\usepackage{enumerate}
\usepackage{natbib}
\usepackage{nicefrac}
\usepackage{subfigure}
\usepackage[colorlinks, citecolor=blue]{hyperref}

\graphicspath{{figures/}}

\begin{document}

\title[Shear thinning fluid \textcolor{black}{flow} through a pipe]{\textcolor{black}{Linear and nonlinear instabilities} in highly shear thinning fluid \textcolor{black}{flow} through a pipe}

\author{
  Xuerao He,\aff{1}
  Kengo Deguchi,\aff{1}
  Runjie Song\aff{1}
  \and Hugh M. Blackburn\aff{2}
}
\shortauthor{He, Deguchi, Song \& Blackburn}

\affiliation{
   \aff{1} School of Mathematics, Monash University, Clayton, VIC 3800, Australia\\[\affilskip]
   \aff{2} Department of Mechanical and Aerospace Engineering, Monash University, Clayton, VIC 3800, Australia
}

\maketitle

\begin{abstract}
Shear-thinning fluids flowing through pipes are crucial in many practical applications, yet many unresolved problems remain regarding their turbulent transition. Using highly robust numerical tools for the Carreau-Yasuda model, we discovered that linear instability can arise when the power-law index falls below 0.35. This inelastic non-axisymmetric instability can universally arise in generalised Newtonian fluids that extend the power-law model. The viscosity ratio from infinite to zero shear rate can significantly impact instability, even if it is small. \textcolor{black}{Two branches of finite-amplitude travelling wave solutions bifurcate subcritically from the linear critical point. The solutions exhibit sublaminar drag reduction, a phenomenon not possible in the Newtonian case.}


\end{abstract}

\section{Introduction}

Research on non-Newtonian fluids is vital for a wide range of applications, including polymer processing, food production, and biomedical engineering \citep{bird2002transport}.
\textcolor{black}{There are \textcolor{black}{four} broadly related classes of shear-thinning fluids: dominantly shear-thinning fluids (e.g., xanthan gum), \textcolor{black}{shear-thinning and yield stress fluids (e.g., Carbopol)}, shear-thinning and viscoelastic fluids (e.g., dilute polymer solutions), and strongly shear-thinning viscoelastic fluids (e.g., polymer melts).} 

Over the years, numerous models have been developed to explain complex behaviour of non-Newtonian fluids.
Among them, our interest lies in a \textcolor{black}{generalised Newtonian fluids (GNF)}, where the shear stress is instantaneously a function of shear rate, thus precluding viscoelastic effects (i.e. Weissenberg number is identically zero). Although GNF models have limitations in explaining experimental results, their simplicity makes them valuable in theoretical and numerical studies.
The simplest GNF model is the power-law fluid model, which assumes that viscosity $\mu$ varies as a power of \textcolor{black}{the strain rate magnitude $\dot{\gamma}$}
\begin{eqnarray}
\mu\propto \dot{\gamma}^{n-1}.\label{powerpower}
\end{eqnarray}
Widely used models, such as the Carreau-Yasuda and Cross models, can be viewed as extensions of (\ref{powerpower}).
When the power-law index $n$ is less than unity, the fluids exhibit shear-thinning behaviour, as observed in substances such as blood (\cite{gijsen1999influence}, 
\cite{boyd2007analysis}) and various industrial fluids (\cite{carreau1979analysis}). However, numerical simulations become extremely challenging as shear-thinning effect intensifies, leaving much of the flow behaviour still poorly understood.

Shear-thinning fluid flow in pipes is one of the most fundamental and practically important cases, motivating numerous experimental investigations (\cite{escudier2005observations,esmael2008transitional,bahrani2014intermittency,charles2024asymmetry}). 
In parallel, extensive numerical simulations have been performed using GNF (\cite{rudman2004turbulent,singh2016importance,gavrilov2017direct,singh2017influence}). Yet, even in such a simple flow configuration, many unsolved problems persist. Even on the fundamental issue of the linear stability of the laminar flow solution, experts remain divided, \textcolor{black}{as will be briefly discussed below.} 

The numerical computation community widely accepts that laminar flow of GNF in a pipe is always linearly stable. This belief stems from the work of \citet{liu2012nonmodal} and \citet{lopez2012pipe},
who performed a linear stability analysis using the Carreau model. 
Somewhat surprisingly, a systematic parameter search for the growth rates of this problem has not yet been reported, presumably because no instabilities have been observed in previous studies.
Consequently, to explain the transition to turbulence, researchers have followed the analyses used for Newtonian pipe flow. For example, \citet{liu2012nonmodal} used the idea of transient growth by \citet{Schmid2001Stability}, while more \textcolor{black}{recently} \citet{plaut2017nonlinear} identified finite amplitude travelling waves analogous to those found in \citet{faisst2003traveling} and \citet{wedin2004exact}. 
Pipe flow of Newtonian fluids is a classic example of shear flow that undergoes subcritical transition, with the amplitude and shape of perturbations that trigger the transition being of great interest to many researchers \citep{avila2023transition}.
Around the transitional Reynolds numbers, it is well known that the flow can be characterised by localised turbulence, called puffs \citep{wygnanski1973transition}. 

Interestingly, in shear-thinning fluids, experiments have observed a transition to asymmetric mean flow profile at values below the critical threshold for puff emergence (see \cite{charles2024asymmetry} and references therein).
While this two-step transition appears typical for strongly shear-thinning fluids, it has not yet been successfully replicated through numerical simulations to the best of the authors' knowledge. 
Recent experimental evidence \citep{picaut2017experimental,wen2017experimental} 
suggests that the emergence of  the asymmetric state might be due to the presence of supercritical bifurcations from the laminar state. Therefore, it is an intriguing question to investigate the types of linear stability that arise in non-Newtonian pipe flows and the nonlinear states that emerge from them.



\textcolor{black}{
Returning to the classification introduced at the beginning of this section, the experiments by \cite{escudier2005observations} 
and \cite{wen2017experimental}
belong to the first class, while \cite{picaut2017experimental} falls into the \textcolor{black}{fourth}. The latter authors reported a critical Weissenberg number, demonstrating that the linear instability originates from viscoelasticity. 
Recently, pipe flow instability has been found in Oldroyd-B fluids, which do not exhibit shear-thinning behaviour, and this has become an active area of research (\cite{garg2018viscoelastic,chaudhary2021linear,dong2022asymptotic}; \cite{sanchez2022understanding}). 
} If instability arises even in the other extreme case, GNF, it would demonstrate that, contrary to common belief, pipe flow can become linearly unstable for a wide class of non-Newtonian fluids.

The next section introduces the mathematical formulation employed in this study, summarising the parameters, the base flow, and the numerical method used for the stability analysis.
\S 3 presents the numerically obtained neutral curves for the idealised parameters. \S 4 then examines parameters relevant to real-world applications. In \S 5, we conduct a bifurcation analysis to identify finite-amplitude states. Finally, we discuss the implications of our results in \S 6.


\section{Formulation of the problem}

\subsection{Governing equations}
Consider an incompressible, shear-thinning fluid through an infinitely long circular pipe. 
We work in cylindrical coordinates $(r,\theta,z)$, where the radial, azimuthal, and axial components of the velocity vector are denoted as $u$, $v$, and $w$, respectively. 
The velocity $\mathbf{u}=[u,v,w](r,\theta,z,t)$ and the pressure $p(r,\theta,z,t)$ are assumed to be governed by the non-dimensional 
\textcolor{black}{Cauchy momentum equation and incompressiblity condition}
\begin{subequations}
	\begin{eqnarray}
		\frac{\partial \mathbf{u}}{\partial t} + (\mathbf{u} \cdot \nabla) \mathbf{u} = -\nabla (p-\frac{q}{Re} z) + \frac{1}{Re}\nabla \cdot (2\mu D),\label{eq.non1} \\
		\nabla \cdot \mathbf{u} = 0,
	\end{eqnarray}
    \end{subequations}
with the strain rate tensor $D=(\nabla\mathbf{u}+(\nabla\mathbf{u})^{\text{T}})/2 $ and  normalised dynamic viscosity $\mu(r,\theta,z,t)$. 
The length scale is the radius of the pipe, $R^* $, the velocity scale is the centre line velocity of the laminar base flow, $U_c^* $, and the pressure scale is $\rho^* U_c^{*2} $, where $ \rho^* $ is the density of the fluid. The scaled constant pressure gradient $q>0$ drives the flow. The no-slip conditions $u=v=w=0$ are imposed on the pipe wall $r=1$. 

We adopt the Carreau-Yasuda 
model \citep{carreau1972rheological,yasuda1981shear} 
\begin{equation}\label{eq:cons}
\mu=\mu_\infty+(1-\mu_\infty)\{1+(\lambda \dot{\gamma})^a\}^{(n-1)/a},
\end{equation}
\textcolor{black}{where the strain rate magnitude $\dot{\gamma}=(2D\negthinspace:\negthinspace D)^{1/2}$ is also called the shear rate in the GNF community.}
The Reynolds number is defined by
	$Re=\rho^* R^*U_c^*/\mu^*_0$,
using the dimensional viscosity at zero shear rate, $\mu^*_0$. 
In (\ref{eq:cons}), $\mu_\infty=\mu_\infty^*/\mu^*_0$, the ratio of viscosities at infinite and zero shear rates,  typically ranges from $ 10^{-3}$ to $ 10^{-4}$ for shear-thinning fluids in experiments \citep{escudier2005observations,escudier2009turbulent}. 
\textcolor{black}{The Carreau number, $\lambda =(U_c^*/R^*)\lambda^*$, can be determined from the `time constant' $\lambda^*$, which represents an inverse shear rate marking the onset of shear thinning.}
When the Yasuda parameter $a=2$, the constitutive relation reduces to that in the Carreau model, and our non-dimensional formulation coincides with that used in \citet{liu2012nonmodal}.

\subsection{Base flow}\label{basesection}
The constant $q$ is determined such that the laminar base flow has a centre line velocity of unity. Substituting $(u,v,w,p)=(0,0,\overline{w}(r),\overline{p}(r))$ into the governing equations, we find that $\overline{w}$ and $q$ can be determined by solving
\begin{subequations}\label{baseuc}
\begin{eqnarray}
	  r^{-1}(r\overline{\mu}\,\overline{w}')'=-q,\qquad \qquad \qquad \label{base} \\\qquad \overline{\mu}(r)=\mu_\infty+(1-\mu_\infty)\{1+|\lambda \overline{w}'|^a\}^{(n-1)/a}, \label{eq.baseu}
\end{eqnarray}
\end{subequations}
subject to the boundary conditions $\overline{w}(1)=0$, $\overline{w}(0)=1$. \textcolor{black}{The primes denote the ordinary differentiation, and overlines indicate laminar base flow quantities.} 
\textcolor{black}{
We solve \eqref{baseuc} for $\overline{w}(r)$ using a numerical method (see Appendix A).  Outcomes agree with the analytical solution involving the Gauss hypergeometric function recently found by \cite{wang22}.
}
For our purposes, numerical computation is more convenient, as it facilitates the easy evaluation of higher-order derivatives of the profile.

\begin{figure}
	\centering
	\subfigure{\label{fig.valispec1}
	\includegraphics[height=5cm]{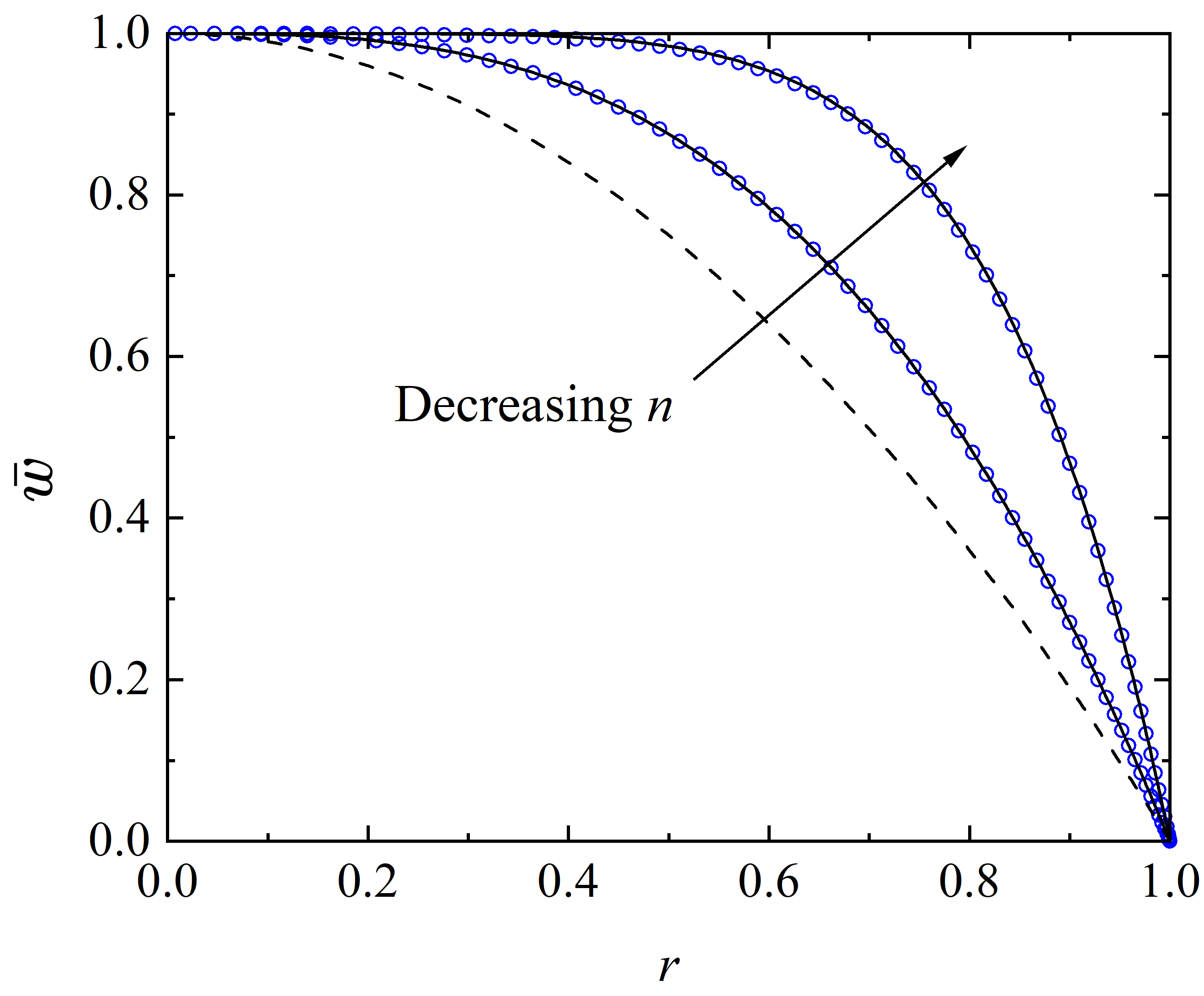}}
    \put(-175,135){(a)}
    \hspace{0pt}
    \subfigure{\label{fig.valispec2}
    \includegraphics[height=5cm]{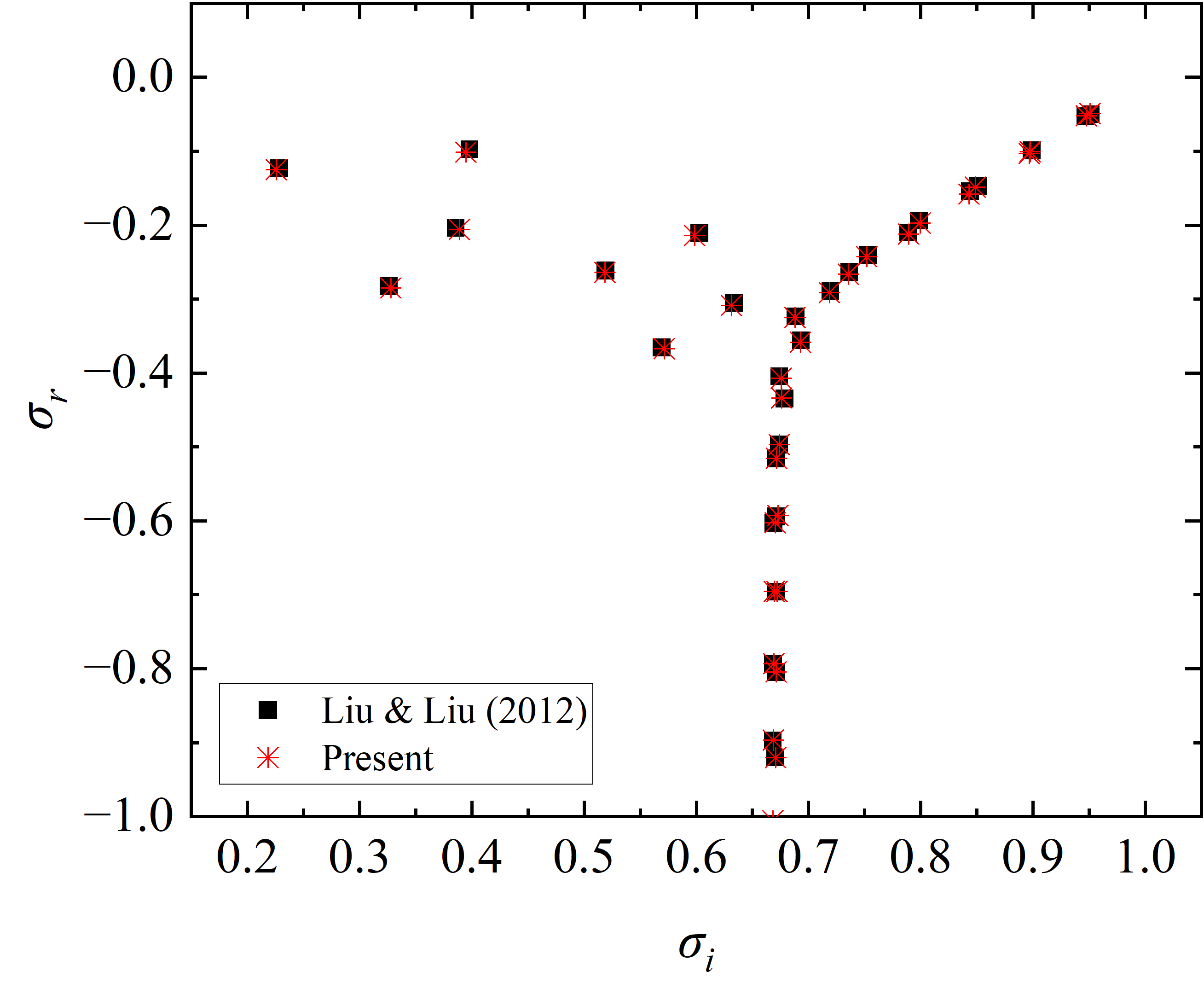}}
    \put(-175,135){(b)}
	\caption{
    (a) Base flow profile for $n=0.5$ and 0.2. The blue symbols are $\overline{w}$ found by the numerical computation of (\ref{base}) with
     $ \mu_\infty = 0 $, $a=2$ and $ \lambda=100 $. 
     The black lines denote the power law approximation $\bar{w}=1-r^{1+1/n}$. The dashed line represents the parabolic base flow profile of the Hagen-Poiseuille flow ($n=1$). 
(b) Complex growth rate at the parameters $ (\mu_\infty,a,n,\lambda,Re) = (0,2,0.8,10,1831.5) $ and the wavenumbers $ k=1 $, $ m=0 $. 
The black squares are taken from  \cite{liu2012nonmodal} for comparison.
}
	\label{fig.base}
\end{figure}

It is common to set $\mu_{\infty}=0$ for simplicity in numerical computations (e.g. \citet{liu2012nonmodal,plaut2017nonlinear}), and we will also examine this idealised case in section 3. In this case, the large $\lambda$ limit of the base flow and the viscosity can be approximated by the power law as
\begin{eqnarray}
\overline{w}=(1-r^{1+1/n})+\cdots, \qquad \overline{\mu}\approx (\lambda |\overline{w}'| )^{n-1}+\cdots\label{power}
\end{eqnarray}
except for a small region around the centreline of the pipe where $r=O(\lambda^{-n})$ (see Appendix A). 
\textcolor{black}{Figure \ref{fig.base}a compares the numerical solution of (\ref{baseuc}) for $\lambda=100$ with the power-law approximation. At this value of $\lambda$, the Carreau fluid almost behaves like a power-law fluid.}


\subsection{Parameters used in experiments}
The five flow parameters $(\mu_\infty,a,n,\lambda)$ and $Re$ defined above are useful for theoretical analysis but are not optimal for organising experimental data. 
We first note that to use the Carreau-Yasuda law, the constants $\mu_0^*$, $\mu_\infty^*$, $\lambda^*$, $n$ and $a$ need to be found by fitting experimental data for the specific fluid in question. Since $\lambda^*$ is a constant particular to the fluid, when the Reynolds number varies in the experiments, $\lambda$ is not a constant but rather a quantity proportional to $Re$. Therefore it is more convenient to specify $\Lambda=\lambda/Re=(\lambda^*\mu_0^*)/(\rho^* R^{*2})$ instead of $\lambda$. Note that $\Lambda$ depends on the pipe radius $R^*$.

Another important consideration is the definition of the Reynolds number. In experiments, it is more convenient to fix the flow rate rather than the pressure gradient, and thus the bulk velocity $U^*_b$ is used as the velocity scale. 
For example, \cite{escudier2005observations} and \cite{wen2017experimental} employed the Reynolds number
\begin{equation}
Re_b=\frac{\rho^* (2R^*) U_b^*}{\langle \mu_{\text{wall}}^* \rangle}
\label{reb}
\end{equation}
using the dimensional viscosity at the wall, $\mu_{\text{wall}}^*$. 
Angle brackets denote the average over $\theta$, $z$, and $t$. 
\textcolor{black}{
Recalling the non-dimensionalisation introduced in section 2.1, we have
$\langle \mu_{\text{wall}}^* \rangle=\mu_0^*\langle \mu \rangle|_{r=1}$ and
$U_b^*=(\int^1_0U_c^*\langle w \rangle r dr)/(\int^1_0 r dr)$. Therefore, the conversion recipe for the two Reynolds numbers can be obtained as
\begin{equation}
Re_b=\frac{4Re}{\langle \mu \rangle |_{r=1}}\int^1_0 \langle w \rangle rdr.
\end{equation}
For the base flow, the ratio $\frac{Re_b}{Re}=\frac{4}{\overline{\mu}|_{r=1}}\int^1_0 \overline{w}rdr$ can be easily computed numerically. 
The power-law fluid approximation (\ref{power}) suggests that for large $\lambda$,
\begin{equation}
\frac{Re_b}{Re}\approx \frac{2n(1+\frac{1}{n})^{2-n} }{(3n+1)}\lambda^{1-n}.\label{Rebconv}
\end{equation}
}




\subsection{Linear stability analysis}

In the linear stability analysis, we assume that the perturbation $[\tilde{u},\tilde{v},\tilde{w},\tilde{p}]=[u,v,w-\overline{w},p-\overline{p}]$ can be written in the form $[\hat{u}(r),\hat{v}(r),\hat{w}(r),\hat{p}(r)]\exp(im\theta+ik(z-ct))+\text{c.c.}$, where c.c. is the complex conjugate. For given flow parameters, the azimuthal wavenumber $m$, and the axial wavenumber $k$, we can numerically solve the linearised governing equations for the complex phase speed $c=c_r+ic_i$.
The resultant stability problem is identical to those described in \citet{liu2012nonmodal}, 
with a correction to the obvious typo in their equation (23):
\begin{subequations}\label{stabeq}
	\begin{align}
		&\hat{u}' + \frac{\hat{u}}{r} + \frac{im}{r} \hat{v} + ik \hat{w} = 0, \\
		&\begin{aligned}
			ik(\bar{w}-c)\hat{u} +\hat{p}'  &= \frac{1}{Re} \left\{ \bar{\mu}  \left[ \mathcal{L} \hat{u} - \frac{\hat{u}}{r^2} - \frac{2im}{r^2} \hat{v} \right] \right. \\
			&\qquad \qquad \qquad \qquad + 2 \bar{\mu}' \hat{u}' + \check{\mu} \left( ik\hat{w}' -k^2 \hat{u} \right) \Bigg\},
		\end{aligned} \\
		&\begin{aligned}
			ik(\bar{w}-c)\hat{v} +\frac{im}{r} \hat{p}&=   \frac{1}{Re} \left\{ \bar{\mu}  \left[ \mathcal{L} \hat{v} - \frac{\hat{v}}{r^2} + \frac{2im}{r^2} \hat{u}\right] \right. \\
			&\qquad \qquad \qquad \qquad + \bar{\mu}' \left( \frac{im}{r} \hat{u} + \hat{v}' - \frac{\hat{v}}{r} \right) \Bigg\},
		\end{aligned} \\
		&\begin{aligned}
			ik(\bar{w}-c)\hat{w} + \bar{w}' \hat{u}+ik \hat{p} &=  \frac{1}{Re} \left\{ \bar{\mu}  \mathcal{L} \hat{w} + (\bar{\mu}'+\check{\mu}'+\frac{\check{\mu}}{r}) \left( \hat{w}' + ik\hat{u} \right) \right. \\
			&\qquad \qquad \qquad \qquad+ \check{\mu} \left( \hat{w}'' + ik\hat{u}' \right) \Bigg\}.
		\end{aligned}
	\end{align}
\end{subequations}
Here, we used the shorthand notations
\begin{eqnarray}
		\check{\mu}(r)&=&(n-1)(1-\mu_\infty)\{1+|\lambda \bar{w}'|^a\}^{(n-1-a)/a}|\lambda \bar{w}'|^a,\label{mucheck}\\
        		\mathcal{L} &=& \frac{\partial^2}{\partial r^2} + \frac{1}{r} \frac{\partial}{\partial r} - \frac{m^2}{r^2} - k^2.
\end{eqnarray}
The boundary conditions become $\hat{u}(1)=\hat{v}(1)=\hat{w}(1)=0$. 

Our numerical code is based on the method described in \citet{deguchi2011bifurcations}, where the poloidal-toroidal potential approach is employed. Spatial discretisation is performed using Fourier-Galerkin and Chebyshev-collocation methods. The radial basis functions follow those used in \citet{deguchi2013swirling}. 
This code has been repeatedly validated by successfully reproducing known travelling waves in Newtonian fluids (see \cite{song2025three}, for example).

Here, we present only the details of the linear stability analysis; for the nonlinear computations, refer to the papers cited above.
In our flow domain, infinitesimally small velocity perturbations are written as $[\tilde{u},\tilde{v},\tilde{w},\tilde{p}]=\nabla \times \nabla \times (\phi \mathbf{e}_r)+\nabla \times (\psi \mathbf{e}_r)$,
where $\phi$ and $\psi$ are poloidal and toroidal potentials, respectively. The two equations for them can be obtained by operating $\mathbf{e}_r\cdot \nabla \times\nabla \times$ and $\mathbf{e}_r\cdot \nabla \times$ on the momentum equations. 
The potentials are then approximated using a finite set of basis functions as follows:
\begin{subequations}
\begin{eqnarray}
\phi=\sum_{l=0}^N a_l \Phi_l^{(m)}(r)\exp(im\theta+ik(z-ct))+\text{c.c.},\\
\psi=\sum_{l=0}^N b_l \Psi_l^{(m)}(r)\exp(im\theta+ik(z-ct))+\text{c.c.},
\end{eqnarray}
\end{subequations}
where c.c. denotes complex conjugate and
\begin{subequations}\label{basisphipsi}
\begin{eqnarray}
\Phi_l^{(m)}(r)&=&
\left \{
\begin{array}{c}
r(1-r^2)^2T_{2l}(r),\qquad \text{if $m=0$},\qquad \qquad \qquad \qquad \\
r(1-r^2)^2T_{2l+1}(r),\qquad \text{if $m$ is odd}, \qquad \qquad \qquad \\
r^3(1-r^2)^2T_{2l}(r),\qquad \text{if $m$ is even and $m\neq 0$},
\end{array}
\right .\\
\Psi_l^{(m)}(r)&=&
\left \{
\begin{array}{c}
r(1-r^2)T_{2l}(r),\qquad \text{if $m$ is even},\qquad  \\
r(1-r^2)T_{2l+1}(r),\qquad \text{if $m$ is odd}.
\end{array}
\right .
\end{eqnarray}
\end{subequations}
Here $T_l(r)$ is the $l$th Chebyshev polynomial.
Evaluating the governing equations at the collocation points $r_k=\cos((k+1)\pi/(2N+4)), k=0,1,\dots,N$, we find that non-trivial coefficients $a_l$ and $b_l$ exist only when \textcolor{black}{the growth rate $\sigma=-ikc$} is an eigenvalue of the resulting algebraic eigenvalue problem. This problem can be solved using LAPACK solvers or Rayleigh quotient iteration scheme. \textcolor{black}{For Newtonian fluids, we verified that the computed eigenvalues match those listed in \cite{schmid1994optimal,meseguer2003linearized} to at least nine decimal places. Shear-thinning effects on the eigenvalues were assessed through comparison with \citet{liu2012nonmodal}; see figure \ref{fig.base}b.}
For most parameters in this paper, the code using $N=200$ achieves excellent convergence without any spurious eigenvalues. 

\section{Linear stability results \textcolor{black}{for $\mu_{\infty}=0$}}

To obtain a general understanding of the stability characteristics of shear-thinning fluids, we first focus on the case where  $a=2$ and $\mu_{\infty}=0$. All the stability results presented in this paper use $m=1$ unless otherwise stated.

\begin{figure}
	\centering
	\includegraphics[height=4.5cm]{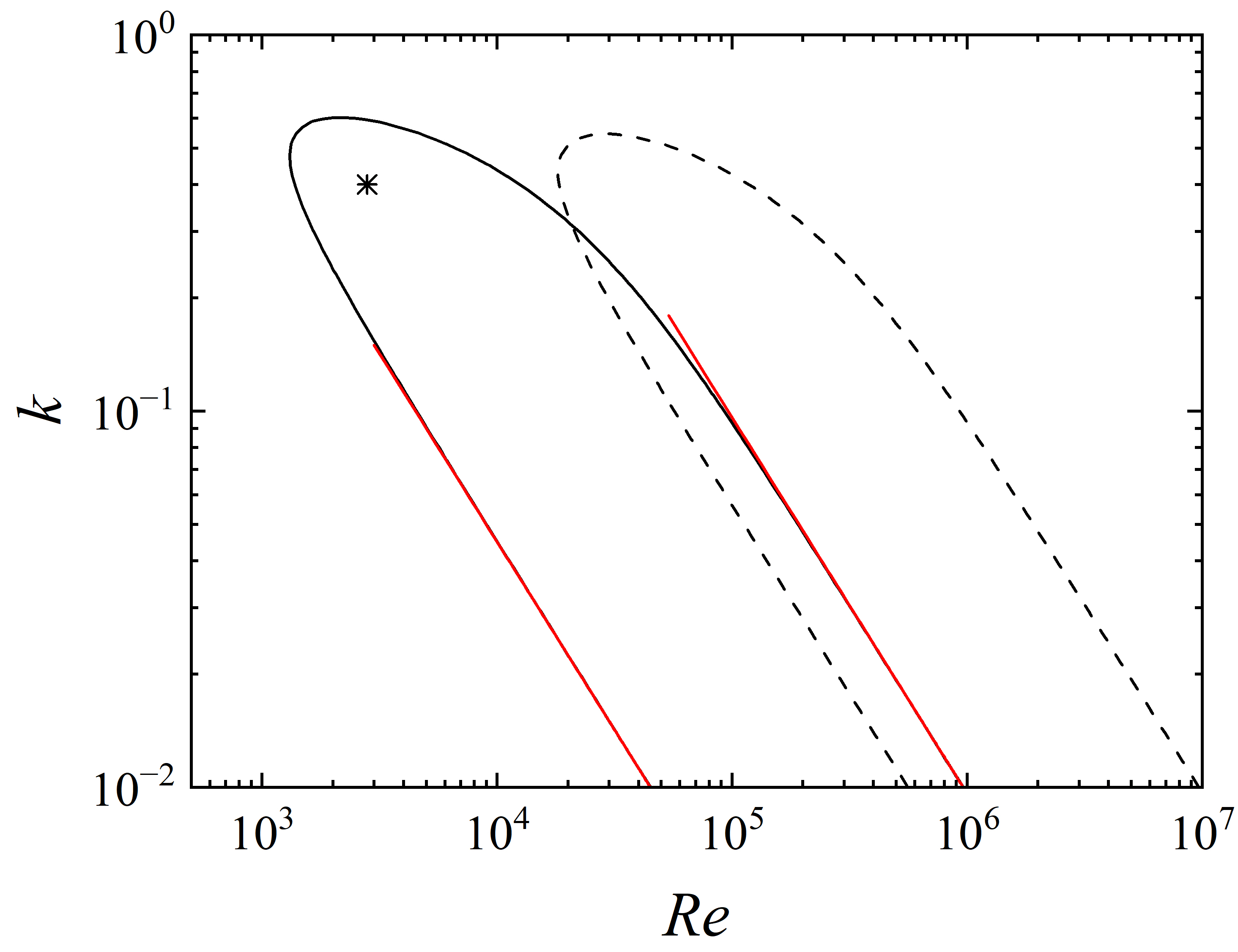}
         \put(-165,120){(a)}\hspace{20pt}
         \includegraphics[height=4.5cm]{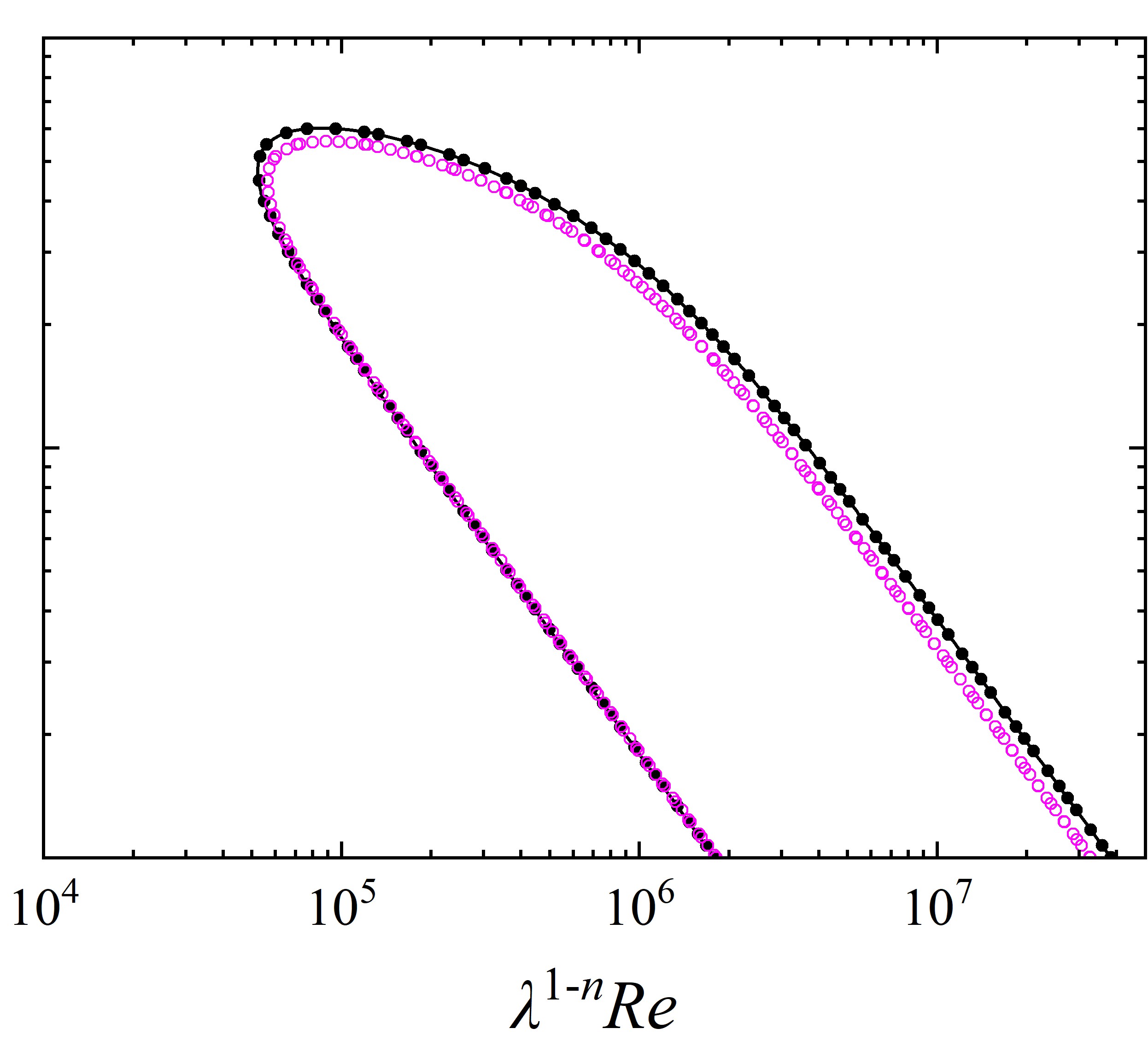}
         \put(-160,120){(b)}
	\caption{Stability results for $n = 0.2$ and $\mu_{\infty}=0$. 
    The Carreau model ((\ref{eq:cons}) with $a=2$) is used except for the magenta open circles. All the instabilities presented in the figures correspond to $m=1$.
    (a) Neutral curves in the $ Re-k $ plane at $\lambda=100$ (solid) and $5$ (dashed). 
    The red lines are the long-wavelength asymptotic results for $\lambda=100$. 
    The star symbol represents the parameter at which the unstable eigenvalue in table \ref{table:ev} is computed. 
    (b) Neutral curves in the $ \lambda^{1-n}Re-k $ plane at $\lambda=50$ (black filled circles) and 100 (solid line). 
The magenta open circles indicate the neutral curve for the Cross model (\ref{Cross}) with $\lambda = 100$. 
    }
    \label{fig.neukRe}
\end{figure}

\begin{table}
	\centering
	\begin{tabular}{cccc}
		   $N$   & $m=1$  & $m=2$  & $m=3$\\
		\hline
        50 & $0.00132743 - 0.0683983i$ & $-0.00267420 - 0.21836877i$ & $-0.00216047 - 0.21928700i$\\
		100 & $0.00127925 - 0.0683962i$ & $-0.00293796 - 0.07518185i$& $-0.00607740 - 0.39262419i$\\
        150 & $0.00127925 - 0.0683962i$& $-0.00293796 - 0.07518185i$ & $-0.00607740 - 0.39262419i$\\
	\end{tabular}
	\caption{The most unstable complex growth rate $\sigma=-ikc$ found at the point marked in figure \ref{fig.neukRe}(a). The parameters are \textcolor{black}{$(a,\mu_{\infty},n,\lambda,Re,k)=(2,0,0.2,100,2800,0.4)$.}
    }
 \label{table:ev}
\end{table}

\begin{figure}
	\centering
	\includegraphics[height=3.5cm]{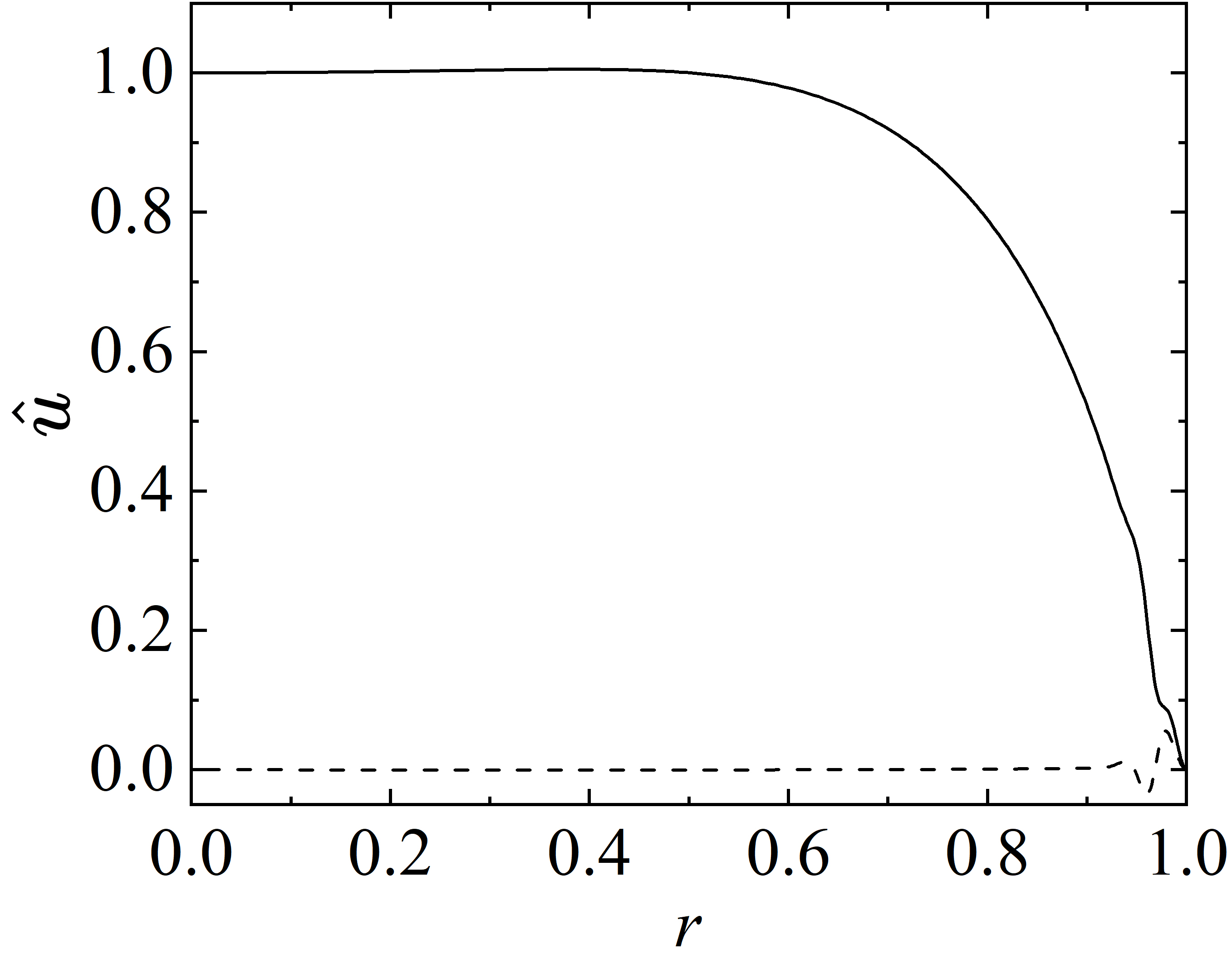}
         \put(-128,92){(a)}\hspace{0pt}
         \includegraphics[height=3.5cm]{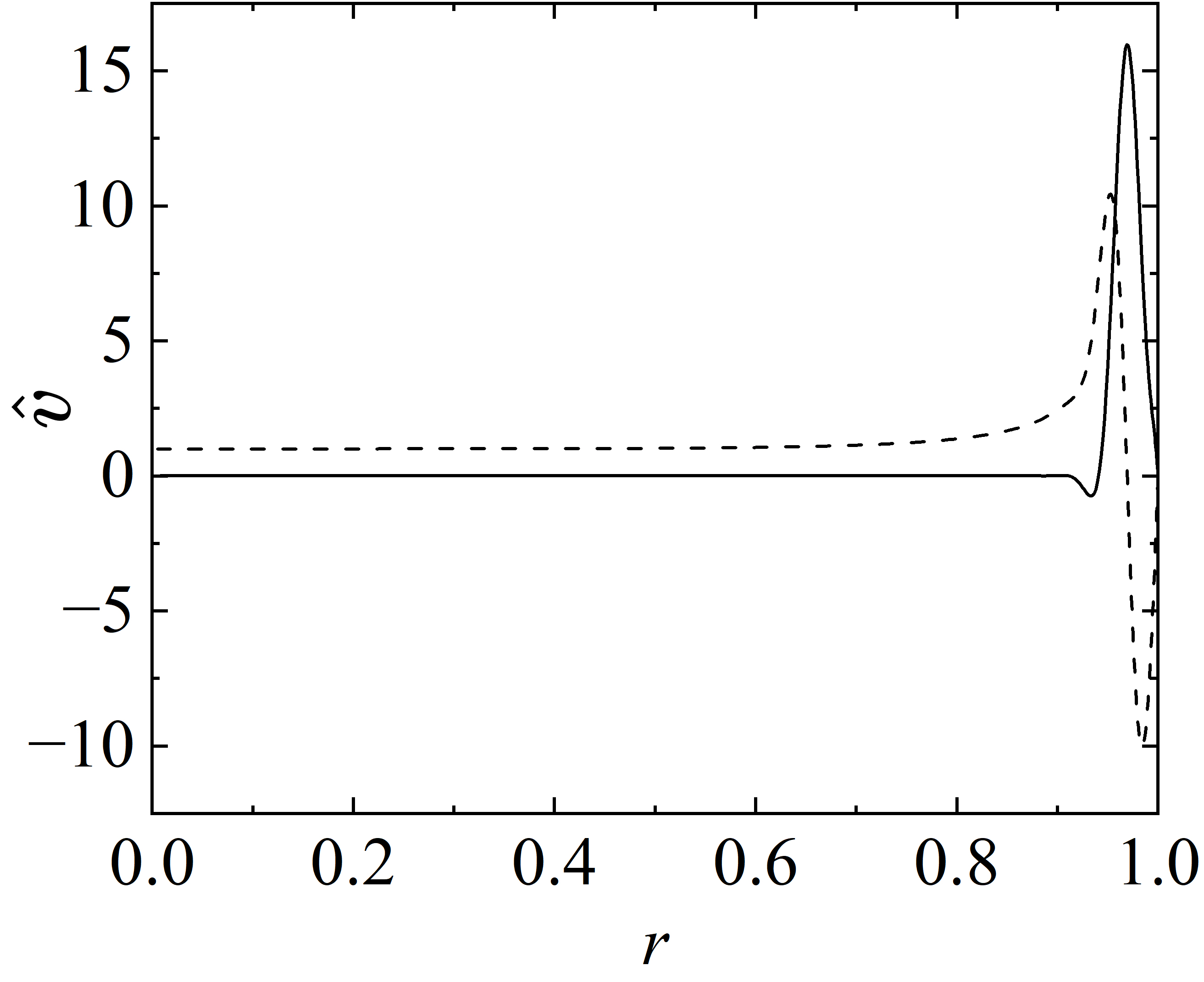}
         \put(-127,92){(b)}\hspace{0pt}
        \includegraphics[height=3.5cm]{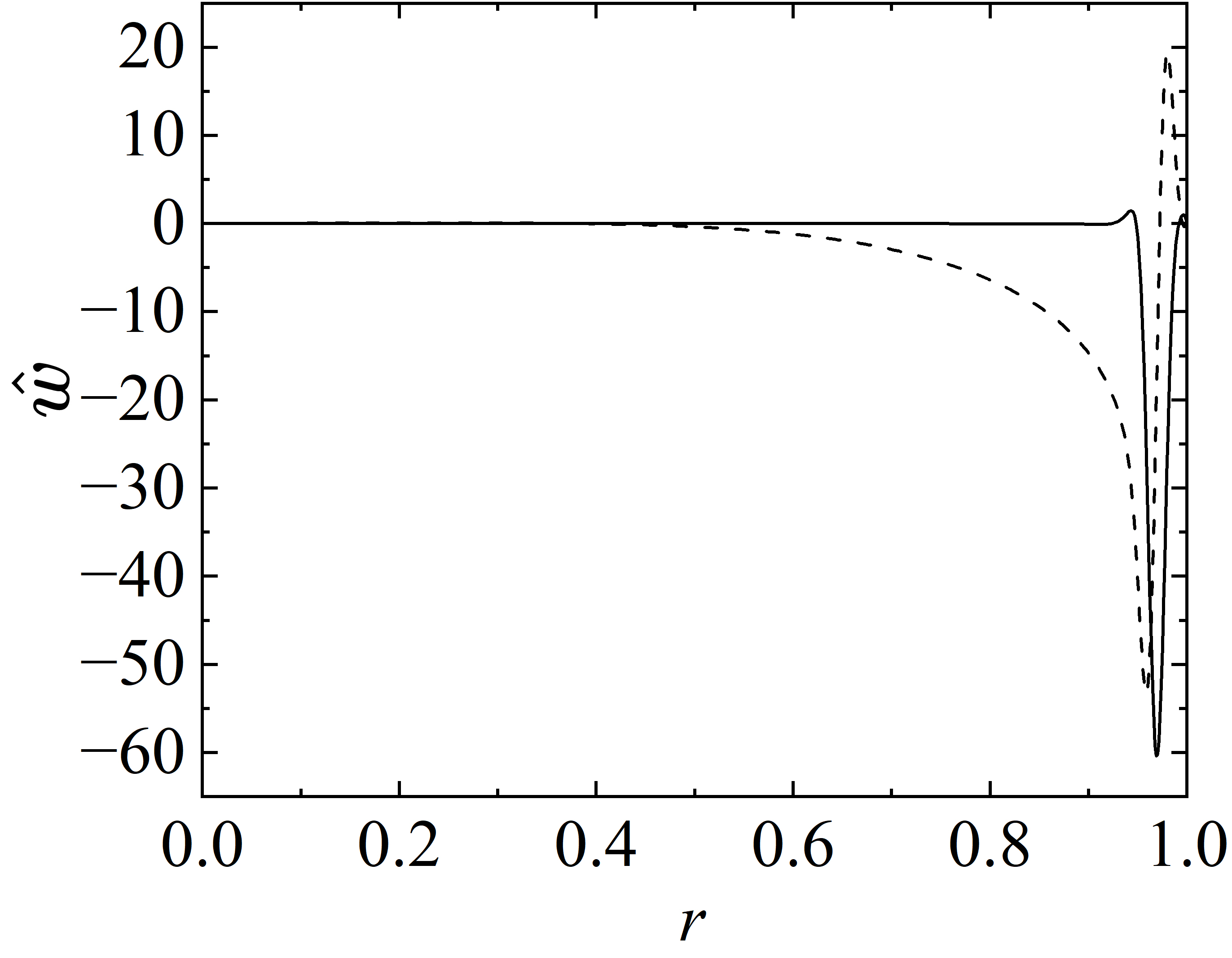}
         \put(-127,92){(c)}
	\caption{Eigenfunction of the unstable mode found at the symbol in figure \ref{fig.neukRe}a. See table \ref{table:ev} for the parameters used. The solid and dashed lines are the real and imaginary parts, respectively.} 
    \label{fig.eigenF}
\end{figure}

\begin{figure}
	\centering
	\label{fig.neucurves}
		\includegraphics[width=9cm]{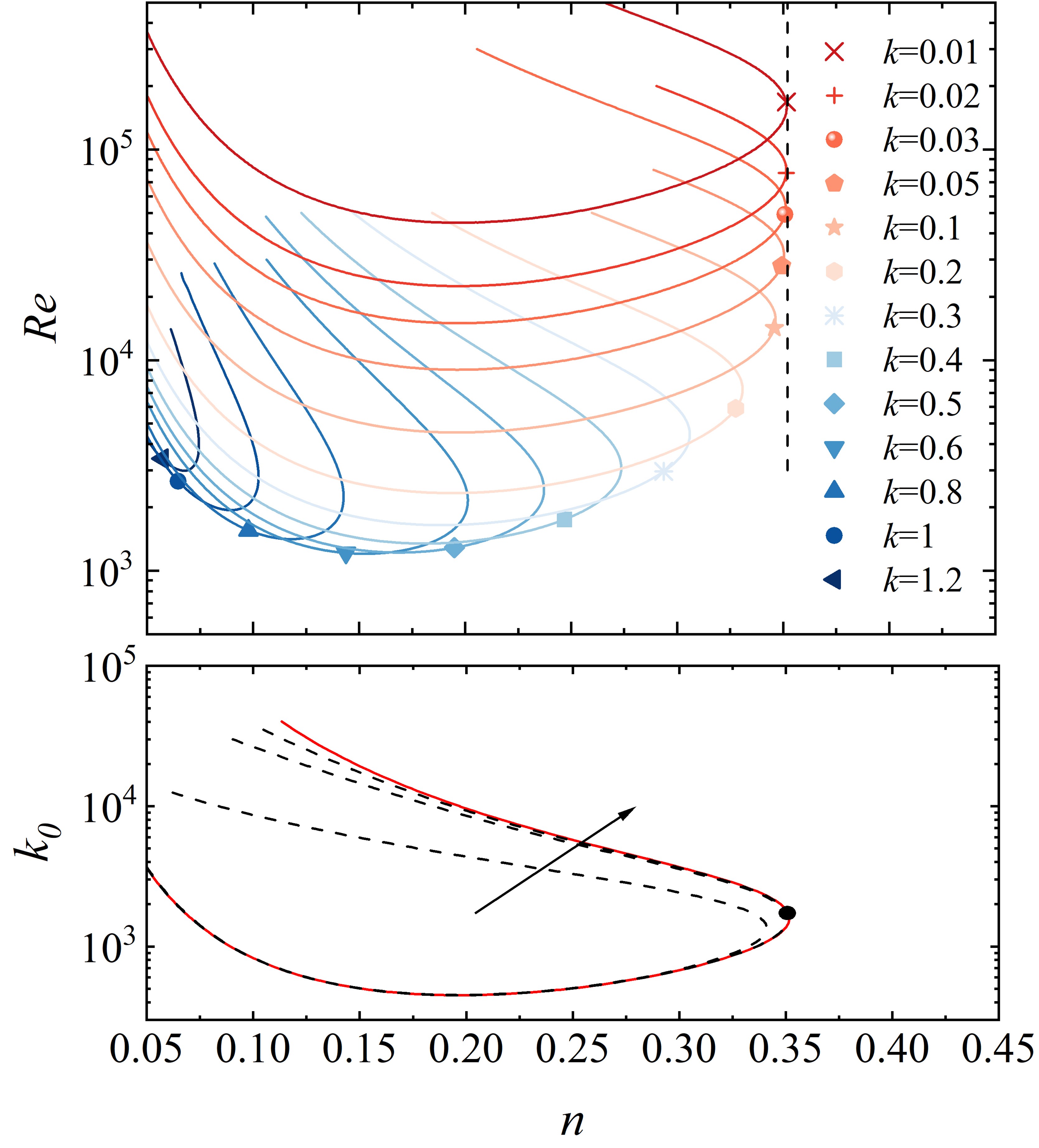}
	\put(-250,280){(a)}
    \put(-95,140){\large Stable}
	\put(-250,120){(b)}
    \put(-115,90){\large Increasing $Re$}
	\caption{
The dependence of the stability results on the power-law index $n$.
	(a) Neutral curves in the $Re$--$n$ plane for $ \lambda=100 $. The dashed line indicates the cutoff value of $n$ for the unstable region. (b) 
	Neutral curves in the $k_0$--$n$ plane at $ \lambda=100 $, where $k_0=kRe$. The black dashed curves are the results for $Re=10^4, 5\times 10^4$, and $10^5$. The red solid curve is the long-wavelength asymptotic result. The circle indicates the point used in figure \ref{fig.asymtotic}a.
	}
	\label{fig.neutral}
\end{figure}

Figure \ref{fig.neukRe}a shows the stability diagram in the $Re$--$k$ plane for $n=0.2$. At this value of $n$, instability does indeed occur for $m=1$. The dashed and solid lines are the neutral curves for $\lambda=5$ and 100, respectively. The region enclosed by the curve is unstable. For example, at $\lambda=100$, an unstable mode is obtained at a point marked by the symbol. The resolution test for this mode is presented in table \ref{table:ev}, confirming that the typically chosen value of $N=200$ is more than sufficient. 
\textcolor{black}{Figure \ref{fig.eigenF} illustrates the corresponding eigenfunction, which is also well-converged.}
For other values of $m$, no instability is observed.


The unstable mode occurs universally in flows that can be approximated by power law fluids. Under the  approximation introduced in section \ref{basesection}, the viscosity functions behave like $\overline{\mu}\approx (\lambda |\overline{w}'| )^{n-1}$ and 
$\check{\mu}\approx (n-1)(\lambda |\overline{w}'| )^{n-1}$. Therefore, the effective Reynolds number in the stability problem (\ref{stabeq}) is $\lambda^{1-n}Re$. In fact, when $\lambda$ is sufficiently large, the neutral curves can be better organised using the rescaled Reynolds number $\lambda^{1-n}Re$ (equivalently $Re_b$, see (\ref{Rebconv})), as shown in figure \ref{fig.neukRe}b. \textcolor{black}{The similar results can be obtained for the Cross model
 \begin{equation}
    \mu=\mu_\infty+\frac{1-\mu_\infty}{1+(\lambda\dot{\gamma})^{1-n}}.\label{Cross}
\end{equation}
The open magenta circles in figure \ref{fig.neukRe}b represent the neutral points for this model with $\mu_{\infty}=0,\lambda=100$.} Apart from the viscosity model, the computational method remains the same as in the Carreau fluid cases. The viscosity defined by (\ref{Cross}) also behaves like a power-law fluid when $\lambda$ is sufficiently large. Consequently, the stability results align with the same universal curve as in the Carreau fluid case when the rescaled Reynolds number $\lambda^{1-n}Re$ is used.

The origin of the instability cannot be an inviscid mechanism, as the sufficient condition for stability established by \citet{batchelor1962analysis} is fulfilled; see Appendix B. Therefore, the unstable mode is of the viscous type. Akin to the Tollmien–Schlichting (TS) wave, in figure \ref{fig.eigenF}, the critical layer $r=0.969$, computed from the phase speed $c=0.171$, is located near the wall. 
It appears that the thin critical layer lies inside a thicker one. This second layer is consistent in thickness with the near-wall boundary layer developed by the base flow when $n$ is small (see figure \ref{fig.base}).



In figure \ref{fig.neukRe}, at large Reynolds numbers, the wavenumber $k$ along the neutral curves is inversely proportional to $Re$, suggesting that the limiting case can be described by asymptotic analysis.
The derivation of the leading-order problem, hereafter referred to as the long-wavelength limit problem, is straightforward. It can be obtained by substituting the regular expansion {$[\hat{u},\hat{v},\hat{w},\hat{p}]=[Re^{-1}\hat{u}_0,Re^{-1}\hat{v}_0,\hat{w}_0,Re^{-2}\hat{p}_0]+\cdots$ into the linearised Navier-Stokes equations, keeping $k_0=Re \,k$ and $c$ as $O(1)$, and discarding the small terms: 
\begin{subequations}\label{longwaveeq}
	\begin{align}
		&\hat{u}_0' + \frac{\hat{u}_0}{r} + \frac{im}{r} \hat{v}_0 + ik_0 \hat{w}_0 = 0, \\
		&\begin{aligned}
			ik_0(\bar{w}-c)\hat{u}_0 +\hat{p}_0'  &=   \bar{\mu}  \left[ \mathcal{L}_0 \hat{u}_0 - \frac{\hat{u}_0}{r^2} - \frac{2im}{r^2} \hat{v}_0 \right]+ 2 \bar{\mu}' \hat{u}_0' + \check{\mu} ik_0\hat{w}_0' ,
		\end{aligned} \\
		&\begin{aligned}
			ik_0(\bar{w}-c)\hat{v}_0 +\frac{im}{r} \hat{p}_0&=    \bar{\mu}  \left[ \mathcal{L}_0 \hat{v}_0 - \frac{\hat{v}_0}{r^2} + \frac{2im}{r^2} \hat{u}_0\right]  + \bar{\mu}' \left( \frac{im}{r} \hat{u}_0 + \hat{v}_0' - \frac{\hat{v}_0}{r} \right),
		\end{aligned} \\
		&\begin{aligned}
			ik_0(\bar{w}-c)\hat{w}_0 + \bar{w}' \hat{u}_0 &=  \bar{\mu}  \mathcal{L}_0 \hat{w}_0 + (\bar{\mu}'+\check{\mu}'+\frac{\check{\mu}}{r}) \hat{w}_0' + \check{\mu} \hat{w}_0'' ,
		\end{aligned}
	\end{align}
\end{subequations}
where $\mathcal{L}_0= \partial^2/\partial r^2 + r^{-1}\partial/\partial r - m^2/r^2$. The boundary conditions are $\hat{u}_0(1)=\hat{v}_0(1)=\hat{w}_0(1)=0$, so the basis functions (\ref{basisphipsi}) can be employed. 
The above equations resemble a linearised version of the Prandtl's boundary layer equations, but they apply across the entire flow region.
Using the parameters from figure \ref{fig.neukRe}a with $\lambda=100$, we can solve (\ref{longwaveeq}) numerically and found that there are two rescaled wavenumbers $k_0=449.6$ and $9635.34$ that make the real part of $\sigma_0$ zero for the leading mode. Those values determine the red lines seen in figure \ref{fig.neukRe}a, which gives a good approximation of the neutral curve when $Re$ is large. As this result suggests, the long-wavelength problem serves as a useful tool for investigating the existence of instability.

Of particular interest for practical applications is determining the values of $n$ for which instability occurs.
Figure \ref{fig.neutral}a shows the neutral stability curves in the $n$--$Re$ plane for $\lambda=100$ and various values of $k$. 
The envelope of the curves shown in the figure gives the stability boundary. 
This boundary seems to exhibit a well-defined cutoff value of $n$ at large Reynolds numbers. Approaching the cutoff, the optimum values of $k$ that define the stability boundary decrease. This behaviour of $k$ is typical when a cutoff in shear-flow instability occurs, as first observed in plane Couette-Poiseuille flow by \citet{cowley1985stability}.
Another example can be found in the study of annular Poiseuille flow by \cite{heaton2008linear}, where a long-wavelength limit system similar to (\ref{longwaveeq}) was derived.


The three dashed curves in figure \ref{fig.neutral}b are the neutral curves for $Re=10^4, 5\times 10^4$, and $10^5$. Here, the vertical axis is the rescaled wavenumber $k_0=Re\,k$. As $Re$ increases, the neutral curves asymptote to the red curve, which is computed using the reduced equations (\ref{longwaveeq}). 
The calculation of the turning point of this curve, indicated by the black circle, provides a convenient way for estimating the cutoff value of $n$ for instability.
The value $n\approx 0.35$ found at the black circle, indicated by the vertical dashed line in figure \ref{fig.neutral}a, indeed represents the critical threshold beyond which the instability no longer occurs at $\lambda=100$.



\begin{figure}	\hspace{10mm}
    	\includegraphics[height=5.8cm]{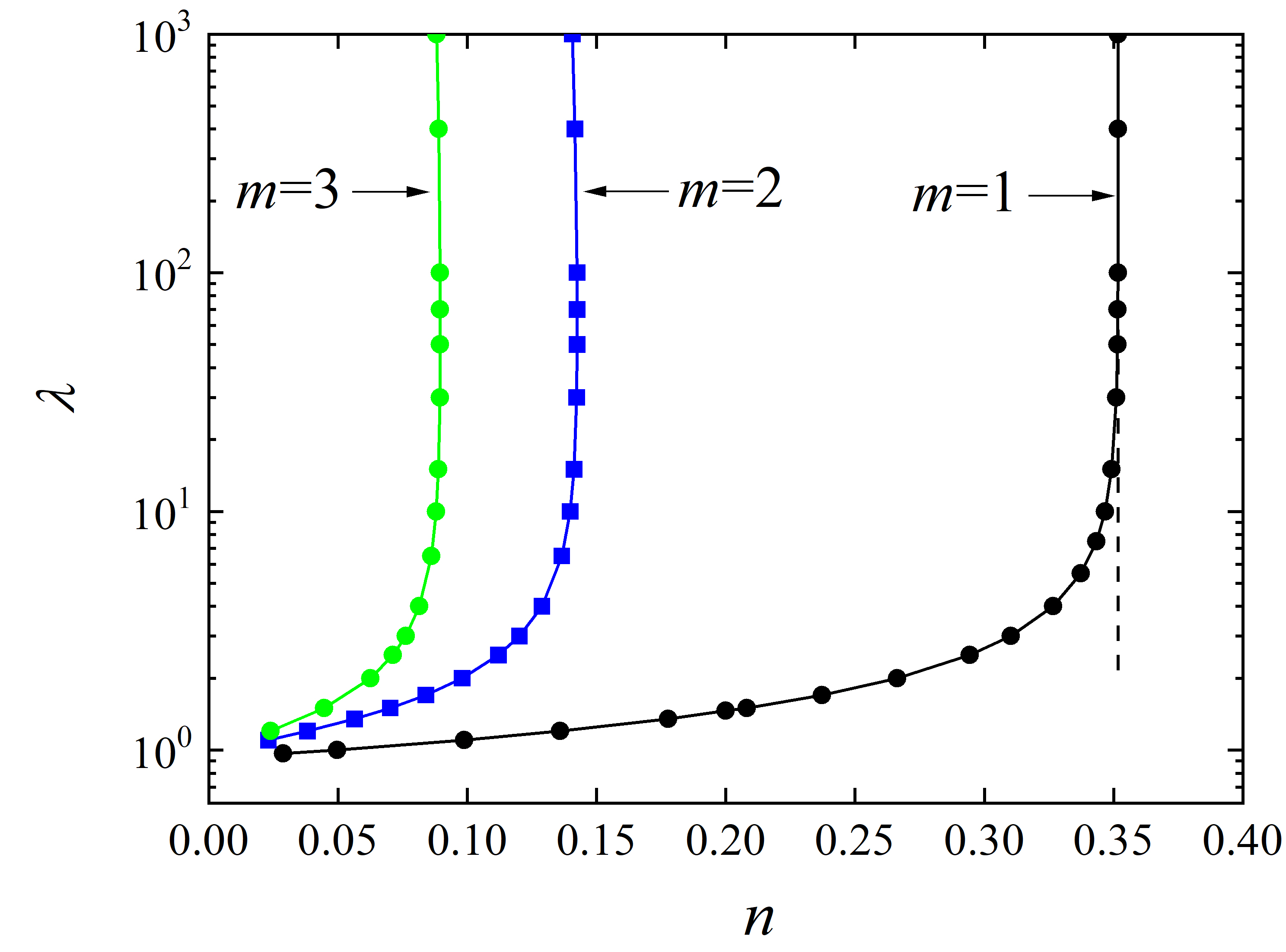}
    \put(-215,155){(a)}
        \put(-40,30){Stable}
    \hspace{10pt}
		\raisebox{6mm}{\includegraphics[height=5cm]{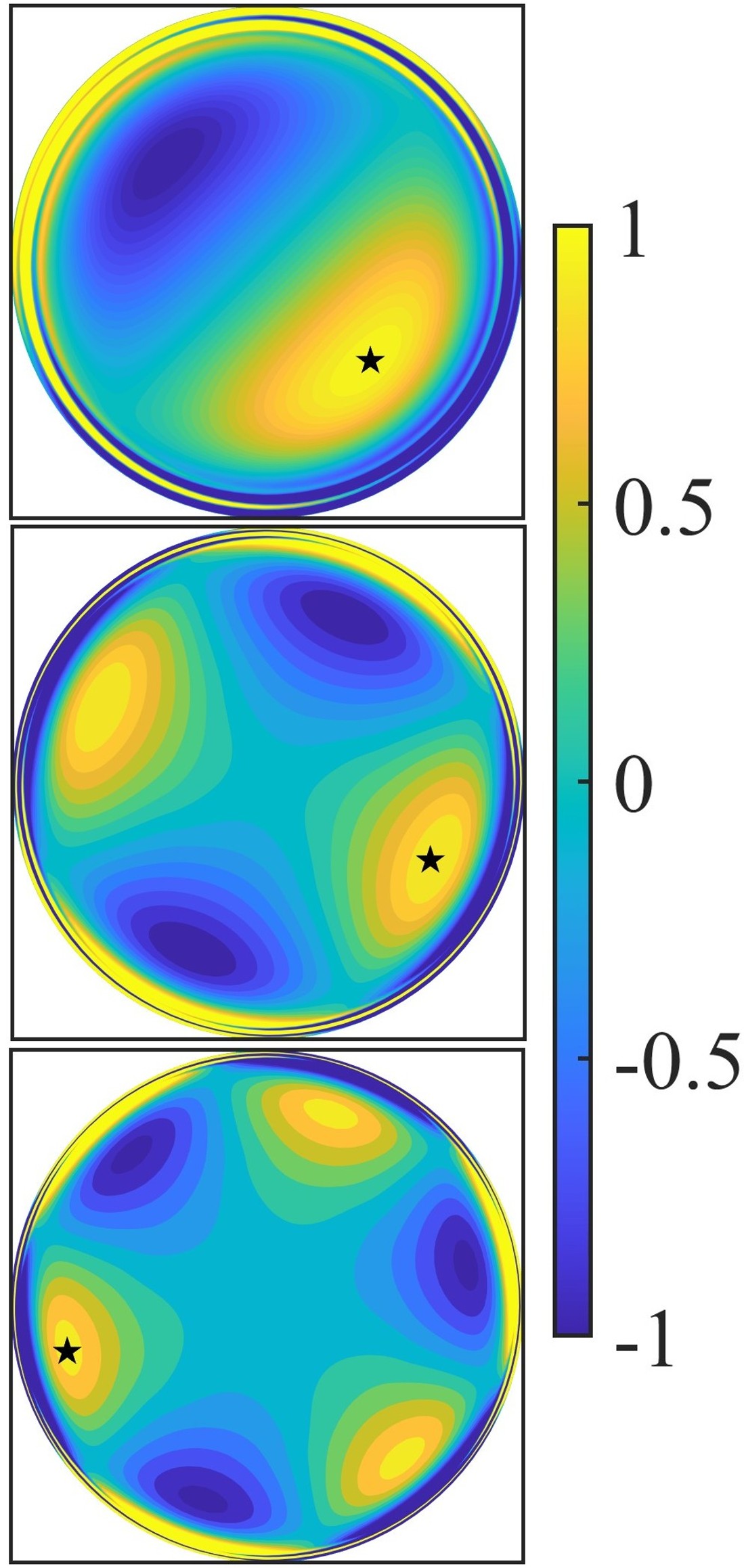}}
    \put(-80,150){(b)}
    \put(-80,105){(c)}
    \put(-80,55){(d)}
	\caption{The stability analysis based on the long-wavelength limit problem (\ref{longwaveeq}).
 (a) Neutral curves in the $n$--$\lambda$ plane with optimised $k_0$.
  (b-d) Streamwise velocity of the neutral eigenfunction in a pipe cross-section, scaled  by its local maximum (marked by a star). 
 The parameters used are (b) $(m,\lambda,n,k_0)=
 (1,1000, 0.3518, 358)$; (c) $(m,\lambda,n,k_0)=
 (2,1000, 0.1407, 163)$ and (d) $(m,\lambda,n,k_0)=
 (3,1000, 0.08816, 167)$.
 }
	\label{fig.asymtotic}
\end{figure}

The long-wavelength limit system (\ref{longwaveeq}) allows us to estimate the thresholds for the existence of the $m=1$ instability for each $\lambda$, as indicated by the black circles in figure \ref{fig.asymtotic}a. The threshold value $n=0.35$ is robust with respect to increases in $\lambda$. This threshold recedes toward smaller values of $n$ as $m$ increases. 
Panels (b), (c), and (d) present a comparison of the neutral modes corresponding to $m=1$, $m=2$, and $m=3$ at $\lambda=1000$.




\section{\textcolor{black}{Linear stability results based on experimentally measured parameters}}

Let us now examine physically meaningful parameters. We first use the long-wavelength limit to narrow down the parameters for which instability may arise. The influence of $\mu_{\infty}$ and $a$, which were held fixed in the previous section, will also be clarified. We then proceed by lifting the long-wavelength assumption to pinpoint the critical Reynolds number.


The open circles in figure \ref{fig.cneb}a present stability results derived from a long-wavelength asymptotic analysis, analogous to those in figure \ref{fig.asymtotic}a but for $(\mu_\infty,a)=(1.116\times10^{-4},2.0)$. These values are taken from the fitting parameters for 7\% aluminum soap (AS); see the first row of table \ref{table:fluid}. 
This fluid has unusually small $\mu_{\infty}$, and its neutral curve closely resembles that for $\mu_{\infty}=0$ (thick grey curve). 
However, noticeable changes emerge as $\mu_{\infty}$ increases.
The filled circles show the stability results for $(\mu_\infty,a)=(1.207\times10^{-3},2.01)$, corresponding to $0.2\% $ polyacrylamide (PAA) used in \cite{escudier2005observations}; see the second row of table \ref{table:fluid}.

\begin{table}
	\centering
	\begin{tabular}{cccccccc}
		   Fluid   & $\mu_0^*$[Pa s]   & $\mu_\infty^*$[Pa s] & $\lambda^*$[s] & $\rho^*$[kg/m$^3$]& $n$ & $a$ & $\mu_{\infty}$\\
		\hline
        $7\%$ AS &89.6&$1\times10^{-2}$&1.41&916&0.2&2.0&1.116$\times10^{-4}$\\
$0.2\%$ PAA& 2.94&$3.55\times10^{-3}$&11.1&1000&0.34&2.01&1.207$\times10^{-3}$\\
Blood          & 0.16&$3.5\times10^{-3}$ &8.2&1000&0.2128&0.64&2.1875$\times 10^{-2}$
	\end{tabular}
	\caption{Carreau–Yasuda model parameters for various fluids.
The first row corresponds to the 7\% aluminum soap (AS) in decalin and m-cresol reported in \cite{myers2005};
the second row to the aqueous solutions of $0.2\% $ polyacrylamide (PAA) listed in
table 1 of \cite{escudier2005observations}; and the
third row to the blood parameters taken from \cite{boyd2007analysis}. 
The values of $\rho^*$ are estimated from the densities of the solvent and solute.}
 \label{table:fluid}
\end{table}


By increasing the parameter $\mu_{\infty}$ of the latter neutral curve to $2.1875\times 10^{-2}$, we obtain the black solid curve in figure \ref{fig.cneb}b. It is evident that, as $\mu_{\infty}$ increases\,---\,even while remaining significantly less than unity\,---\,stabilisation occurs in the large $\lambda$ parameter region. Repeating the discussion in \S 2.2 while keeping $\mu_{\infty}$ clarifies that $\mu_{\infty}=O(\lambda^{n-1})$ is large enough to influence the behaviour of viscosity (\ref{eq.baseu}). In fact, for $\mu_{\infty}\gg O(\lambda^{n-1})$, the viscosity behaves nearly Newtonian, leading to the absence of instability. 
The dashed curves illustrate that a decrease in the value of $a$ also contributes to stabilisation. While the instability exists at $a=1$, they disappear when $a$ is reduced to 0.64. This value, along with $\mu_{\infty} = 2.1875 \times 10^{-2}$ and $n=0.2128$, corresponds to the Carreau-Yasuda parameters for blood as reported in \cite{boyd2007analysis}.

\begin{figure}
	\centering
	\subfigure{\label{fig.criticnexp}
		\includegraphics[height=4.5cm]{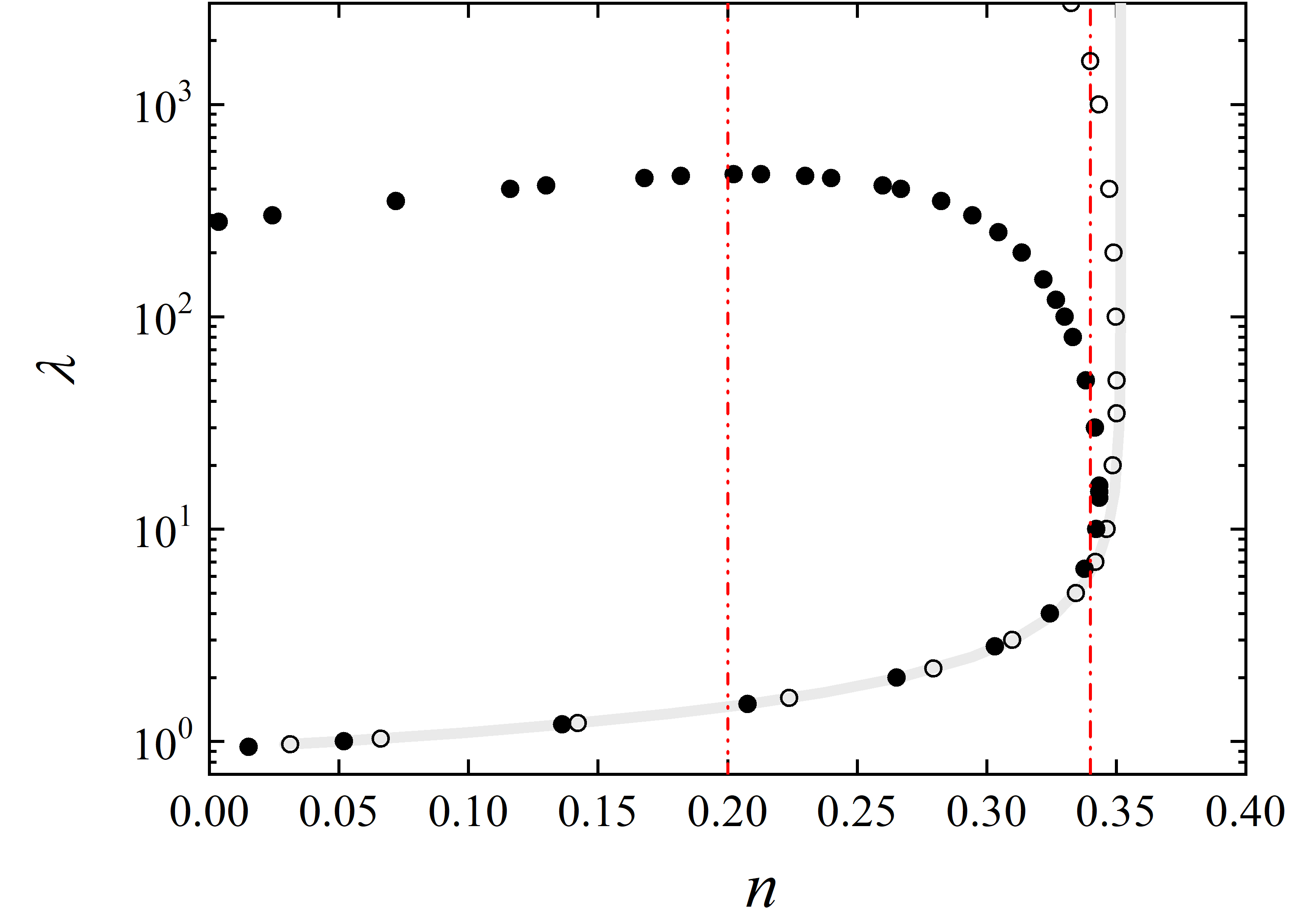}}
	\put(-175,120){(a)}
	\hspace{0pt}
	\subfigure{\label{fig.criticnblood}
		\includegraphics[height=4.5cm]{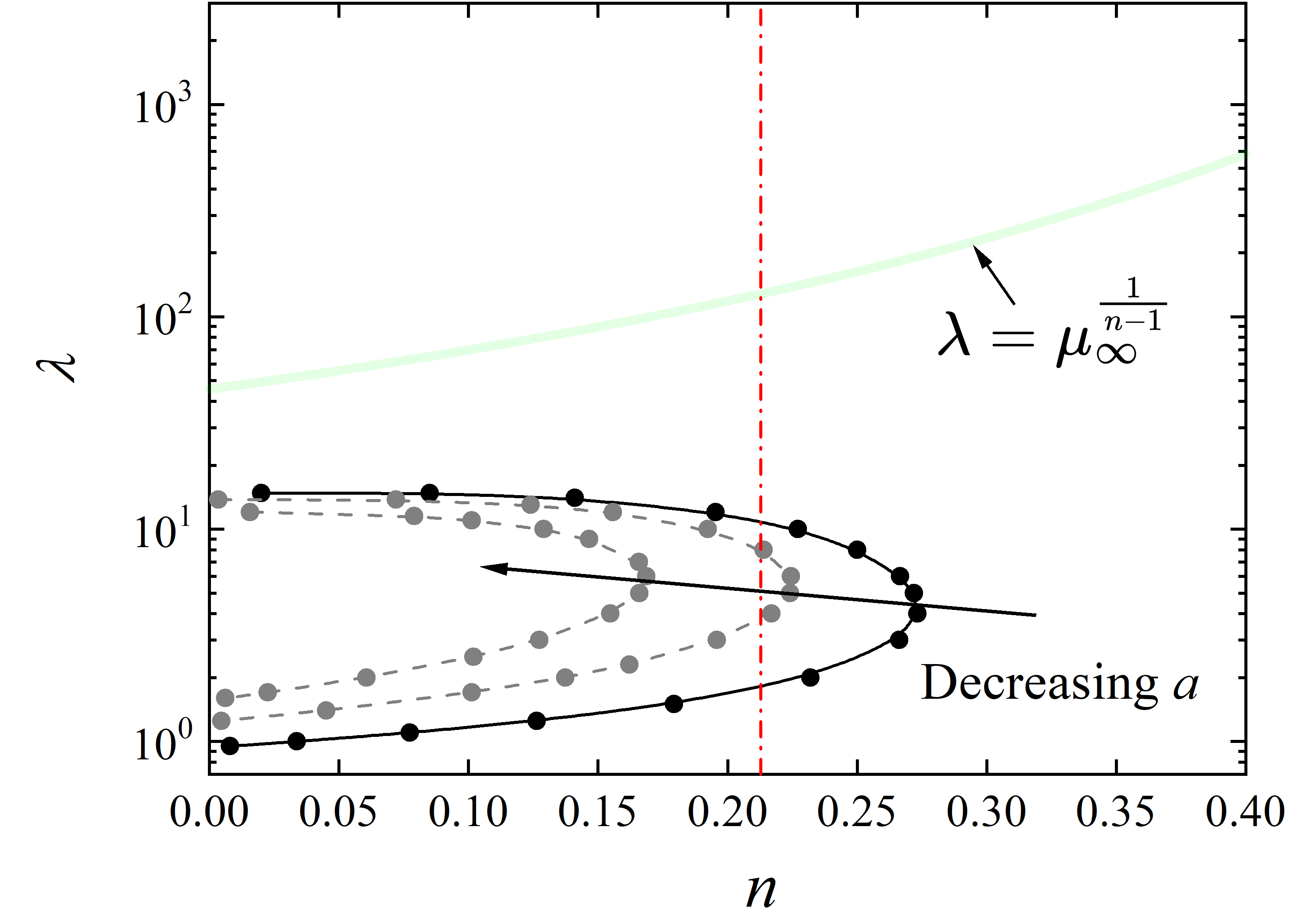}}
	\put(-175,120){(b)}
	\caption{
 Similar plots to figure \ref{fig.asymtotic}a, using different parameters $ (\mu_\infty,a)$.
 (a) The open and filled circles are results for $ (\mu_\infty,a) = (1.116\times10^{-4} , 2.0)$ and $ (\mu_\infty,a) = (1.207\times10^{-3}, 2.01)$, respectively. For comparison, the thick curves with light colour represent the results with $(\mu_\infty,a) =(0,2)$ (same as figure \ref{fig.asymtotic}a). 
 The red dash-dot-dot and dash-dot lines are the value of $n$ for 7\% AS and 0.2\% PAA (see table \ref{table:fluid}).
 (b) The solid curves with symbols are computed using $ (\mu_\infty,a) = (2.1875\times10^{-2},2.01)$. 
 For the two dashed curves with a lighter colour, $a$ is set to 1.3 and 1.0, respectively.
 The red dash-dot line indicates the value of $n$ for blood (see table \ref{table:fluid}).
 %
 }
	\label{fig.cneb}
\end{figure}

It should be noted that while the long-wavelength approximation employed in figure \ref{fig.cneb} is useful for assessing whether instability occurs for certain choices of $k,Re,$ and $\lambda$, identifying the physically relevant critical Reynolds number requires solving the full stability problem. Figure \ref{fig.ReB}a presents the numerical results for the 0.2\% PAA parameters. Here, following the remark in \S 2.3, we fix $\Lambda$ and vary $Re_b$. 
The parameter $\Lambda$ is determined not only by the fluid properties but also by the pipe radius $R^*$, and it has a significant influence on the flow stability. 
To observe instability, $\Lambda$ needs to be as small as $5.5\times10^{-4}$, which corresponds to a pipe radius of approximately $R^*=8$[m]. \textcolor{black}{\cite{escudier2005observations} employed a much smaller radius $R^*=$0.05[m] in their experimental apparatus. The use of this parameter yields $\Lambda = 13.05$, which unfortunately does not exhibit instability.}
Theoretically, this result is not surprising, given that figure \ref{fig.cneb}a indicates the potential for instability only near $\lambda = O(10)$. The predicted critical Reynolds number of $O(1)$, obtained from the relation $\Lambda = \lambda/Re$, is too low. 
\textcolor{black}{In contrast, using 7\% AS allows instability to occur with the experimentally feasible pipe radius of $R^* = 0.05$[m] ($\Lambda=55.17$), as shown by figure \ref{fig.ReB}b.} In all cases, the instability forms an isolated region in the $k$–$Re_b$ plane.
Note that increasing $Re_b$ leads to flow stabilisation, since the value of $\lambda$ eventually exceeds the Newtonian recovery threshold, $\mu_{\infty}^{1/(n-1)}$.

We also carried out stability analyses using the blood parameters. The most practically relevant values of $\Lambda$ are $8.13$ and $345.04$, corresponding to the aorta ($2R^*=2.54\times 10^{-2}$[m]) and brachial artery ($2R^*=3.90\times 10^{-3}$[m]), respectively, as listed in table I of \cite{boyd2007analysis}; however, no instability was detected for either case. Additional tests with even smaller $\Lambda$ values yielded the same result. This outcome is consistent with the long-wavelength limit analysis shown in figure \ref{fig.cneb}b, where no sign of instability was observed for $n=0.2128, a=0.64$.

\begin{figure}
	\centering
	\subfigure{\label{fig.Rebgr}
		\includegraphics[height=4.5cm]{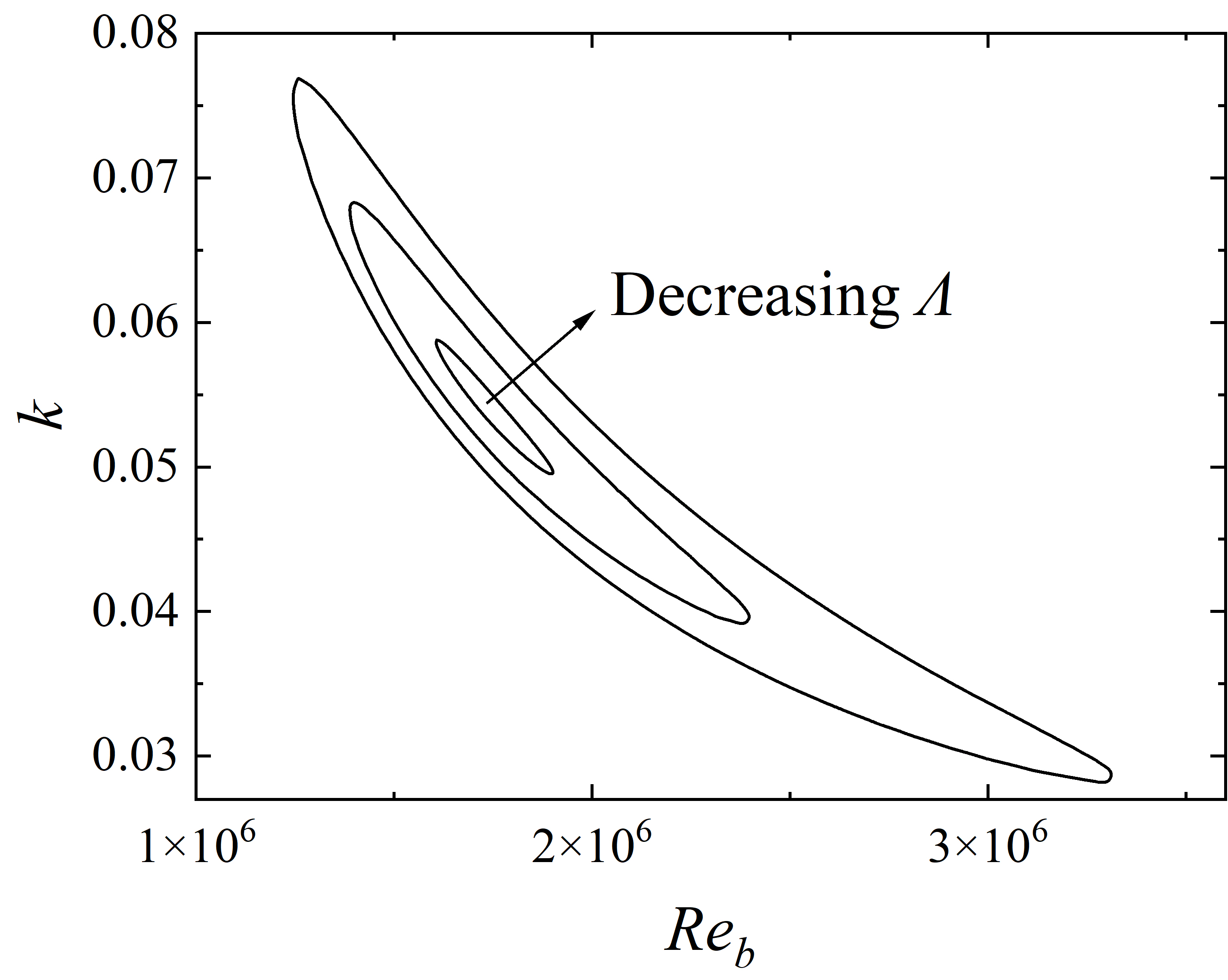}}
	\put(-165,120){(a)}
	\hspace{0pt}
	\subfigure{\label{fig.RebgrB}
		\includegraphics[height=4.5cm]{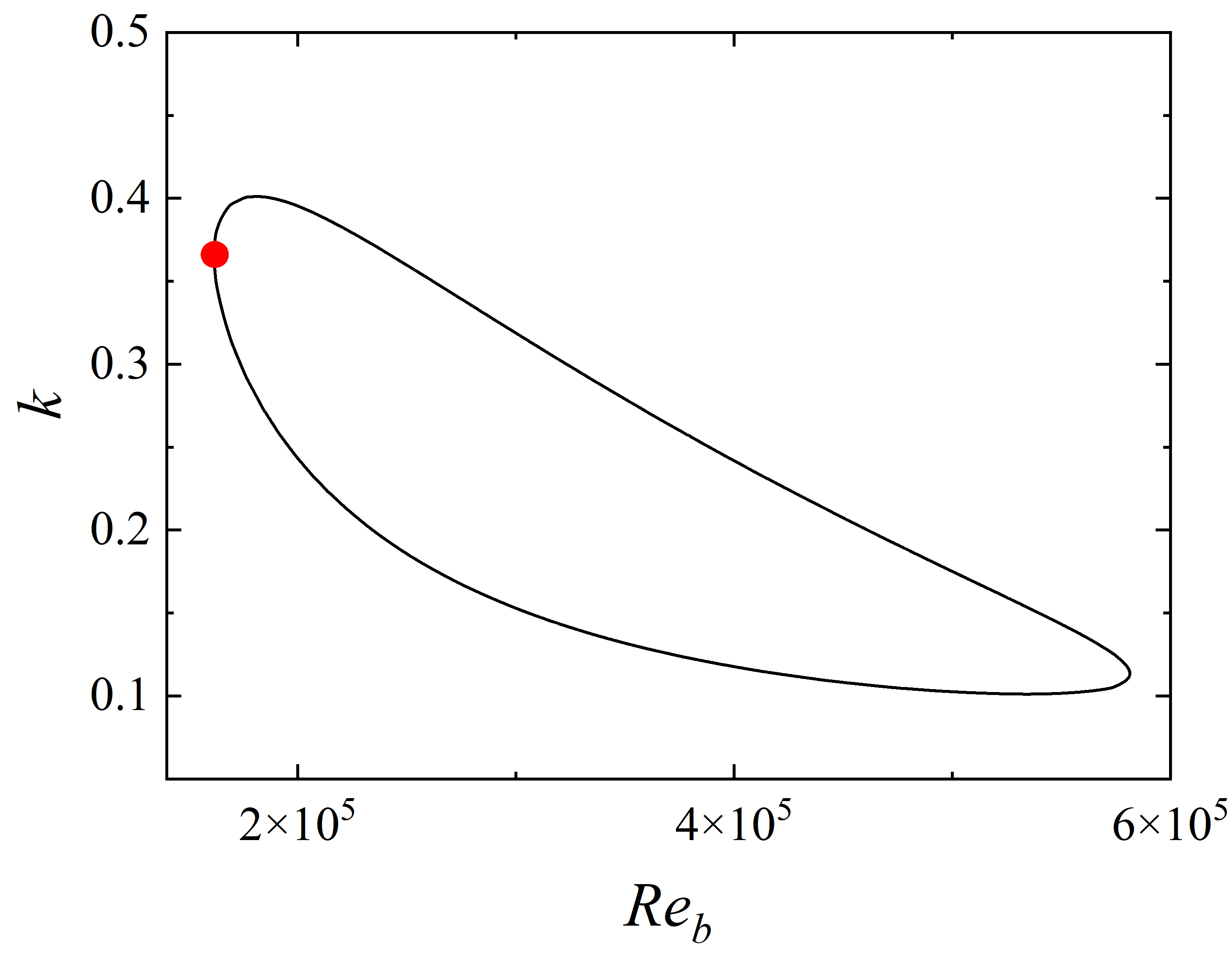}}
	\put(-170,120){(b)}
	\caption{Neutral curves in the $k-Re_b$ plane. Parameter values are listed in table \ref{table:fluid}. (a) 0.2\% PAA. The values of $\Lambda$ used are $5.5\times10^{-4}$, $5\times10^{-4}$, and $4\times10^{-4}$. 
 (b) $7\%$ AS with $\Lambda=55.17$. The red bullet indicates the critical point
 $(k_c,Re_{b,c})\approx (0.366,1.621\times 10^5)$.
 }
	\label{fig.ReB}
\end{figure}

\section{Bifurcation analysis}

Bifurcation theory suggests that finite-amplitude travelling wave solutions emerge from the linear critical points identified so far. 
This section targets the critical point indicated by the red bullet in figure \ref{fig.ReB}b.

A naive approach to obtaining nonlinear solutions is to integrate the governing equations in time near the critical parameter, hoping for convergence to a finite amplitude equilibrium state. Thus, we first attempted direct numerical simulation near the linear critical parameters using the spectral element solver, \textit{Semtex} \citep{blackburn2019semtex,blackburn2025semtex}. However, for $n<0.5$, numerical instability arose unless a very small time step was chosen. Similar computational challenges for small $n$ have also been reported in \cite{plaut2017nonlinear} and \cite{wen2017experimental}.
Furthermore, the wavelength of the perturbation that yields the critical value $Re_{b,c}$ in figure \ref{fig.ReB}b is fairly long, while high resolution is required in the radial direction. Therefore, we conclude that obtaining a travelling wave state through direct numerical simulation is not feasible.

Our computation here instead utilises Newton's method implemented in the code by \citet{deguchi2011bifurcations}. This code employs an analytically derived Jacobian matrix, which eliminates the need for time integration and thus avoids the aforementioned numerical instability. However, deriving the analytic Jacobian matrix is a cumbersome task as $\mu$ depends on the perturbation. \textcolor{black}{This motivates us to consider the Taylor expansion of the viscosity
\begin{eqnarray}\label{taylor}
\mu=\overline{\mu}+\frac{\check{\mu}}{\overline{w}'}\frac{\partial \tilde{w}}{\partial r}+\cdots,
\end{eqnarray}
which is valid when the perturbation $\tilde{\mathbf{u}}=(u,v,w-\overline{w})$ is smaller than the base flow $\overline{w}$. In our numerical computations, we adopt an approximation where the expansion is truncated at the second term. This approximation is justified for bifurcating solutions near the linear critical point and ensures accurate reproduction of weakly nonlinear analysis results, such as the Landau coefficient. Retaining Fourier modes up to the fourth harmonic is sufficient, since we only investigate the vicinity of the bifurcation point.}


\begin{figure}
	  \centering
		\includegraphics[height=15cm]{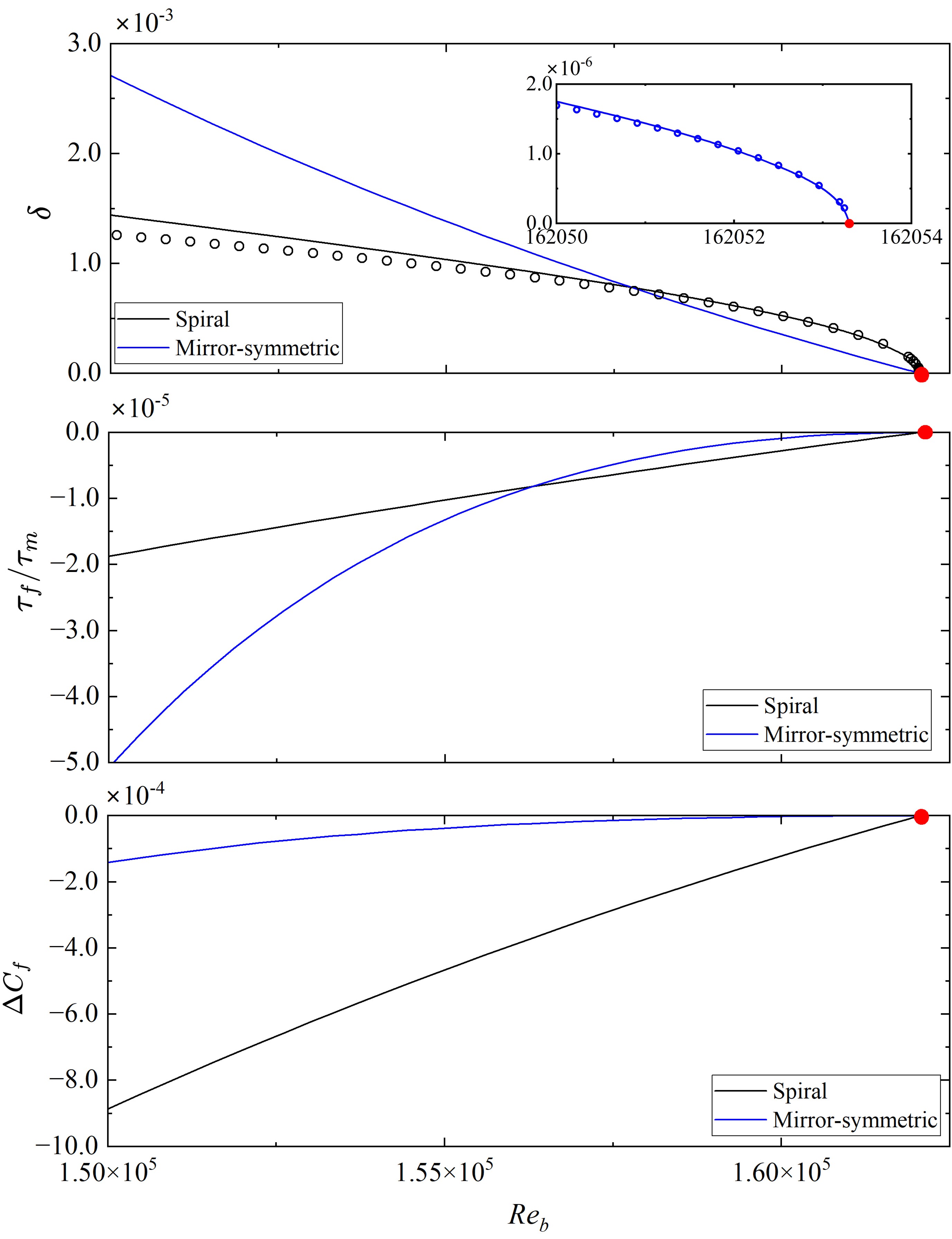}
     \put(-320,410){(a)}
	\put(-320,270){(b)}
	\put(-320,145){(c)}
		\caption{
  Bifurcation analysis from the neutral point in figure \ref{fig.ReB}b. $7\%$ AS with $\Lambda=55.17$.
    The axial wavenumber is fixed at $k=0.366$.
  (a) Bifurcation diagram based on the energy norm of the velocity perturbation, $\delta$.
  (b) Same results, shown the ratio of the shear stress from fluctuations to that of the mean flow. (c) Same results, shown in terms of the friction coefficient. 
  %
  }
  \label{fig.nonlinear}
\end{figure}

\begin{figure}
\centering
		\includegraphics[height=7cm]{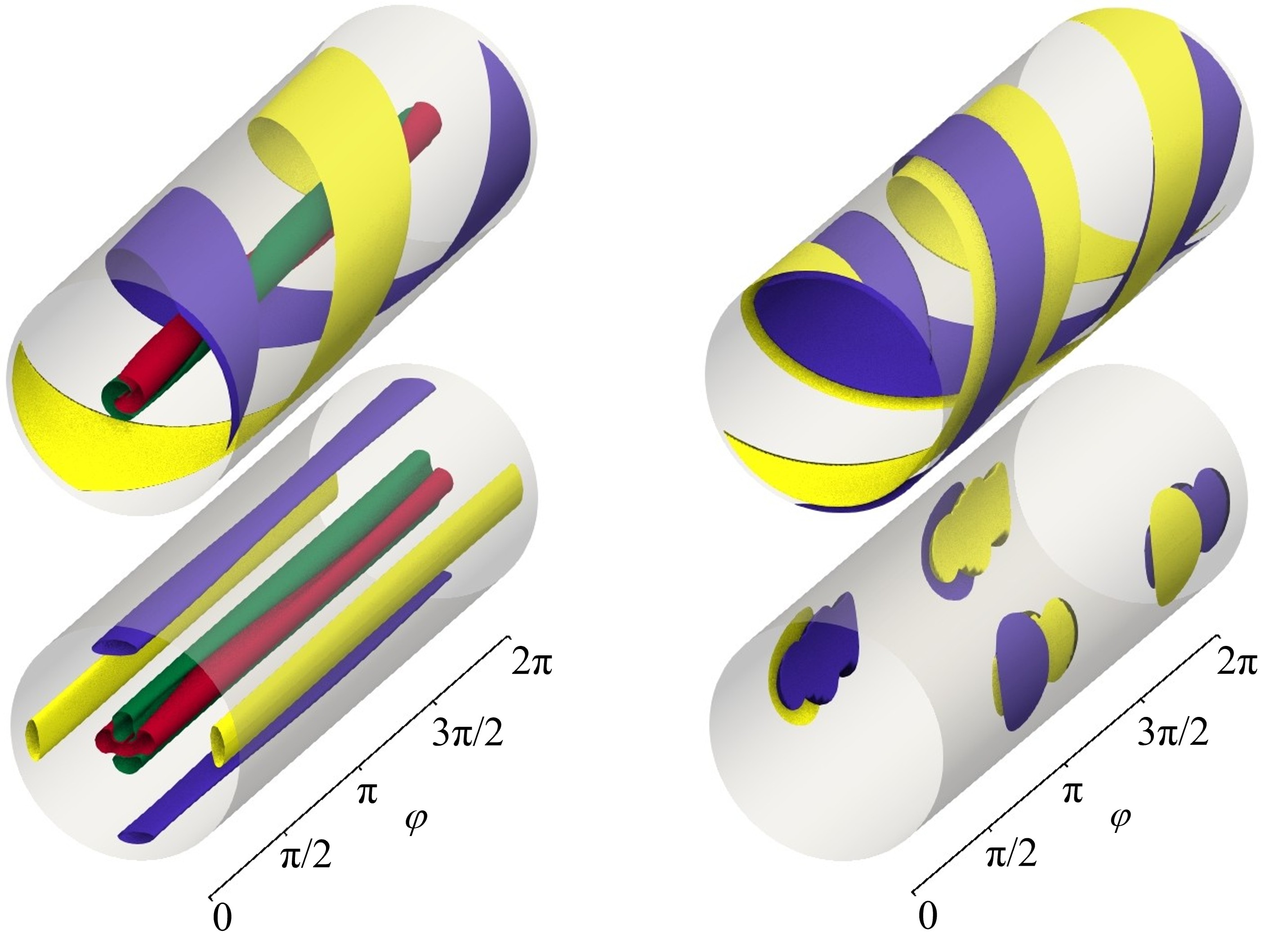}
        	\put(-280,190){(a)}
	      \put(-130,190){(b)}
            \put(-280,90){(c)}
	      \put(-130,90){(d)}
		\caption{
        Perturbation flow field of the solutions at $Re_b=1.52\times10^5$. 
The phase is defined by $\varphi=k(z-ct)$, where $c$ is the phase speed of the travelling wave. 
Panel (a) shows the streamwise velocity $\tilde{u}$ of the spiral solution. The yellow/blue surfaces depict the positive/negative isosurfaces at 88\% of the maximum magnitude. 
The red/green surfaces represent the positive/negative isosurfaces at 10\% of the maximum magnitude for the viscosity variation $\mu-\bar{\mu}$. 
Panel (b) is the vorticity $\tilde{\omega}$ of the spiral solution, with the isosurfaces plotted at 88\% of the maximum magnitude.
Panels (c) and (d) show plots similar to panels (a) and (b), but for the mirror-symmetric solution.
  }
  \label{fig.nonlinearvel}
\end{figure}

\begin{figure}
	\centering
	\subfigure{\label{fig.detamuspi}
		\includegraphics[height=5.5cm]{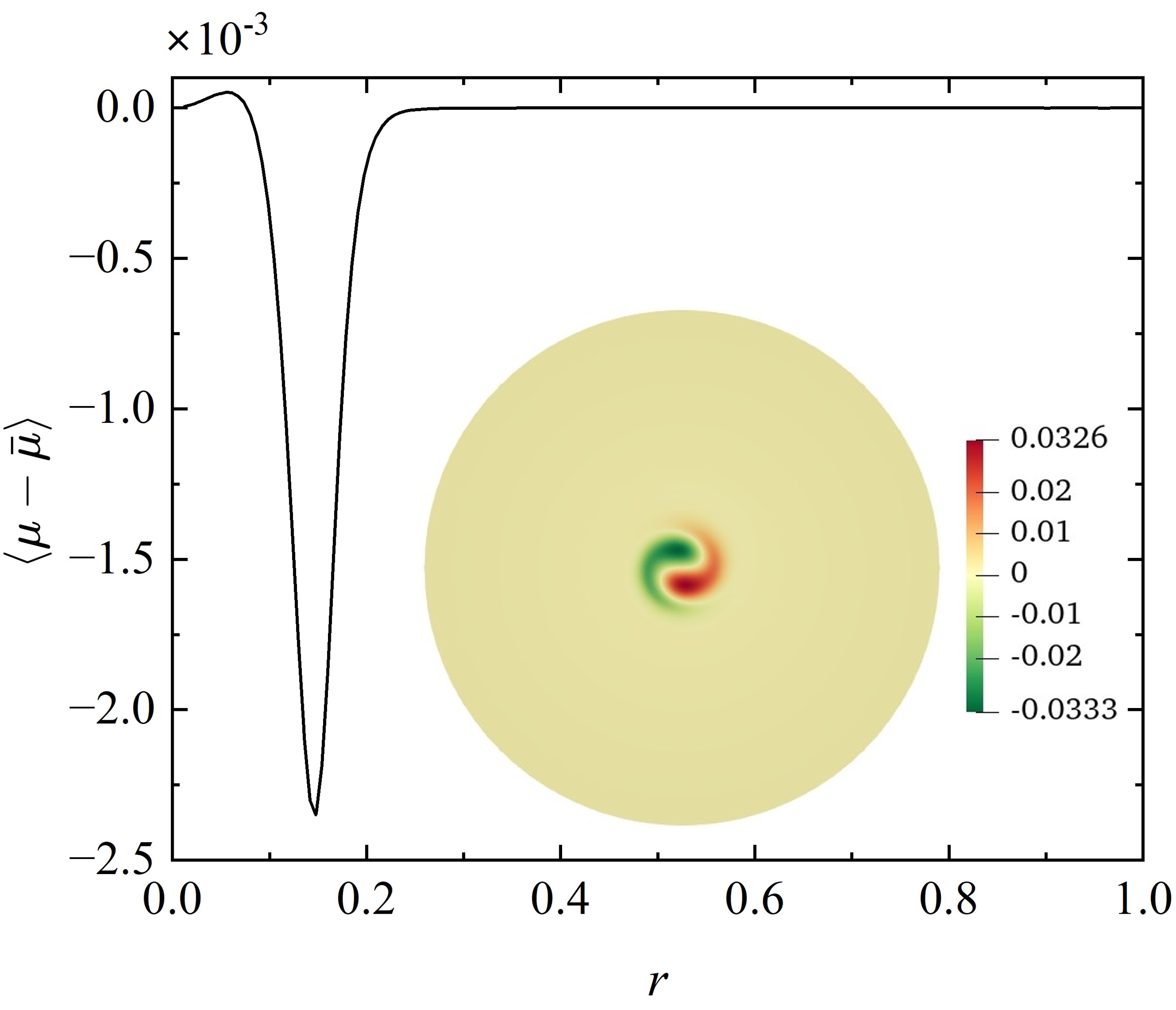}
		}
	\put(-180,145){(a)}
	\hspace{0pt}
    	\subfigure{\label{fig.detamurib}
		\includegraphics[height=5.5cm]{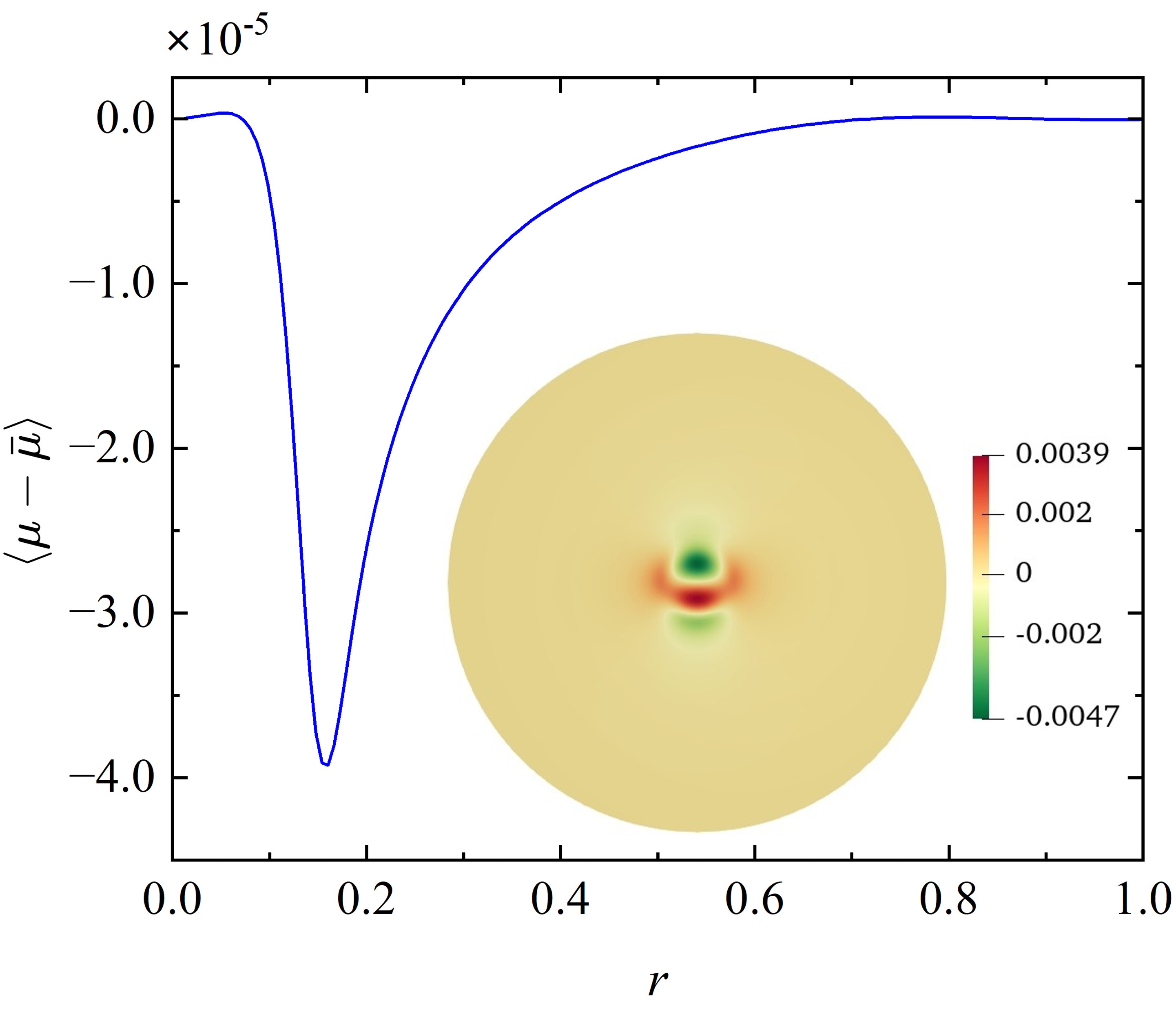}
		}
        \put(-180,145){(b)}
	\caption{The curves indicate the value of $\langle\mu-\bar{\mu}\rangle$ as a function of $r$ for (a) spiral solution; (b) mirror-symmetric solution. $Re_b=1.52\times10^5$. The insets denote colour maps for $\mu-\bar{\mu}$ at $\varphi=0$.
    }
    \label{fig.mumu}
\end{figure}


The amplitude equations derived from the weakly nonlinear analysis suggest the existence of two types of solutions. This is confirmed by our computations, as shown in figure \ref{fig.nonlinear}a. 
We start the Newton's method by using a single neutral eigenfunction as an initial guess. 
With an appropriate choice of amplitude, the iterations converge, resulting in the black curve in figure \ref{fig.nonlinear}a.  
In this bifurcation diagram, we measured the solutions by the normalised energy norm of the velocity perturbation,
\begin{equation}
\delta=\left[\int^1_0\langle \tilde{u}^2+\tilde{v}^2+\tilde{w}^2\rangle rdr~\Big/\int^1_0rdr\right]^{1/2}.
\end{equation}
The angle bracket denotes $\theta$--$z$ average. 
The structure observed in the isosurface of $\tilde{u}$, shown in figure \ref{fig.nonlinearvel}a, displays characteristics that prompt us to refer to this solution as the `spiral solution'. The helical invariance of the solution clearly arises from the linear neutral eigenfunction with $m=1$. 
This invariance is also evident in the streamwise vorticity, $\tilde{\omega}=r^{-1}\partial_{\theta}\tilde{w}-\partial_z \tilde{v}$ (figure \ref{fig.nonlinearvel}b).

The symmetry of the system indicates that when a helical neutral mode with a specific pitch exists, there is always another helical mode with the opposite pitch. Using a superposition of the symmetric pair of neutral modes as the initial condition for Newton's method leads to convergence to a `mirror-symmetric solution' (the blue curve in figure \ref{fig.nonlinear}a). 
The symmetry of the solution is evident from the isosurfaces shown in figures \ref{fig.nonlinearvel}c and d. 

As seen in figure  \ref{fig.nonlinear}a, for both solution branches, the bifurcation is subcritical. 
Given that $\delta \ll 1$, the approximation based on the Taylor expansion (\ref{taylor}) is considered reasonable.
The circles show a square-root fit, indicating the weakly nonlinear regime. The weakly nonlinear approximation for the mirror-symmetric solution is only valid very close to the bifurcation point, as shown in the inset. Note that the control parameter in our numerical code is $Re$. Thus to draw figure  \ref{fig.nonlinear}a, we first compute the values of $Re_b(Re)$ and $\delta(Re)$ for each solution, and then plot the parametric curve in the $Re_b$--$\delta$ plane.

The numerically obtained solutions exhibit features that are not observed in the travelling waves of Newtonian fluids. In Newtonian pipe flow, the pressure gradient $q/Re$  can be readily determined from the mean flow alone. However, for GNF, it is also necessary to account for the fluctuating components. This can be found by taking the spatial average of the $z$-component of the momentum equation (\ref{eq.non1}):
\begin{eqnarray}
\frac{1}{2}\frac{q}{Re}=-\frac{1}{Re}\langle \mu \partial_r w \rangle|_{r=1}=\tau_m+\tau_f.
\end{eqnarray}
On the right-hand side of this global momentum balance, 
$\tau_m=-\frac{1}{Re}\langle \mu \rangle \langle w \rangle' |_{r=1}$ represents the shear stress from the mean flow, while 
$\tau_f=-\frac{1}{Re}\langle (\mu-\langle \mu \rangle)\partial_r(w- \langle w \rangle)\rangle |_{r=1}$ corresponds to the contribution from fluctuations, which is absent in the case of Newtonian fluids. As seen in figure \ref{fig.nonlinear}b, the signs of $\tau_m$ and $\tau_f$ are opposite in our solution.

Figure \ref{fig.nonlinear}c shows the variation of
\begin{eqnarray}
\Delta C_f=\frac{C_f-\overline{C}_f}{\overline{C}_f}.\label{delCf}
\end{eqnarray}
Here, $C_f$ is the friction factor, which is commonly defined by
\begin{eqnarray}
C_f=\frac{4R^*\Delta p^*}{(U_b^*)^2}=\frac{4q}{Re U_b^2},\label{cfdefinition}
\end{eqnarray}
using the dimensional pressure gradient driving the flow, $\Delta p^*$, and the non-dimensional bulk velocity, $U_b=U_b^*/U_c^*=2\int^1_0\langle w\rangle rdr$. Following the above remark, the definition uses the pressure gradient instead of wall shear. In (\ref{delCf}), $\overline{C}_f$ denotes the friction factor of the laminar flow with the same $Re_b$ as the travelling wave solution.
Therefore, figure \ref{fig.nonlinear}c suggests that, surprisingly, the resistance experienced by the pipe decreases from the laminar state. 

Such `sublaminar drag' is widely known to be impossible in Newtonian fluids (see \cite{marusic2008minimum} and \cite{fukagata2009lower}, for example). 
An analysis of the energy equation in Appendix C reveals that the following inequality holds.
\begin{eqnarray}
0\leq\frac{8}{U_b^2Re_b \overline{C}_f \langle \mu \rangle|_{r=1}}\int^1_0\langle(\overline{\mu}-\mu)  D\negthinspace:\negthinspace D\rangle rdr
+\left . \left (\frac{\langle \mu \rangle-\overline{\mu} }{\langle \mu \rangle}\right)\right |_{r=1}+\Delta C_f.\label{energyCf}
\end{eqnarray}
The second term on the right hand side is simply a consequence of using the wall viscosity in the definition of $Re_b$ (see (\ref{reb})), and is therefore of little physical significance. 
The first term in (\ref{energyCf}) is central to the possibility of  sublaminar drag reduction in GNF. 
The importance of variation of viscosity, $\mu-\overline{\mu}$, can also be found in the generalised FIK identity shown in Appendix D. Compared with the Newtonian version by \cite{fukagata2002}, an additional term depending on the viscosity variation appears (see (\ref{FIK2})). 

The viscosity variation $\mu-\overline{\mu}$ is concentrated near the pipe centre; see figure \ref{fig.nonlinearvel}ac. 
This may seem counterintuitive at first. However, this behaviour is consistent with (\ref{taylor}), where the second term is proportional to $\check{\mu}$. Recall that for large $\lambda$, the approximation $\check{\mu}\approx (n-1)(\lambda |\overline{w}'| )^{n-1}=O(\lambda^{n-1})\ll 1$ holds almost everywhere except near the pipe centre. 
The $\mu-\overline{\mu}$ field exhibits a complex structure, and it is not immediately clear how it contributes to the integral in (\ref{energyCf}). Nevertheless, its $\theta$--$z$ average, $\langle\mu-\bar{\mu}\rangle=\langle\mu\rangle-\bar{\mu}$, shown in figure \ref{fig.mumu}, at least plays a role in reducing drag, as the figure suggests that $\int^1_0(\overline{\mu}-\langle \mu \rangle)  \langle D\negthinspace:\negthinspace D\rangle rdr>0$.

\textcolor{black}{Since the definition of $Re_b$ depends on $\langle \mu_{\text{wall}}^* \rangle$ (see (\ref{reb})), one may wonder whether sublaminar drag can be observed using other choices of reference viscosity. We performed comparisons using both the average viscosity and $\mu_0^*$, and found that the conclusion remains unchanged. Drag reduction can be more directly confirmed by keeping the pressure gradient fixed and comparing the resulting flow rates.}



\section{Conclusions and discussion}

We found that the laminar state of 
\textcolor{black}{GNF}, described by the Carreau-Yasuda model and 
flowing through a pipe, can become linearly unstable. 
\textcolor{black}{This instability universally occurs in fluids that can be approximated by power-law models, including the Cross law. 
} 
The unstable modes generate non-axisymmetric vortices near the pipe wall. This structure differs entirely from the axisymmetric `center mode' instability identified by \cite{garg2018viscoelastic} in Oldroyd-B pipe flow. We have proven that inviscid instability does not occur in the shear-thinning Carreau–Yasuda model (Appendix B). Therefore, the origin of the new instability is viscous in nature, like TS waves.
However, the high Reynolds number limit of our mode differs from that of the TS waves; it corresponds to a long-wavelength limit, with the wavelength scaling with the Reynolds number.

The appearance of instability requires very strong shear-thinning behaviour, with a power-law index $n<0.35$. 
Moreover, the effect of $\mu_{\infty}$ on stability is not negligible. For example, although blood is characterised by a small power-law index $n$, its relatively large $\mu_{\infty}$ makes instability unlikely due to shear-thinning effect alone. However, instability is indeed possible under certain experimentally feasible conditions. One such example is the 7\% aluminum soap in decalin and m-cresol flowing through a pipe of radius $R^*=$0.05[m].
Using the Carreau–Yasuda parameters reported in \cite{myers2005}, we found that a spiral perturbation with $m=1$ and axial wavelength 0.86[m] becomes unstable. From the linear critical point, a nonlinear spiral travelling wave bifurcates subcritically. In addition, the superposition of two linear spiral perturbations with opposite pitch gives rise to a mirror-symmetric solution, which also bifurcates subcritically. 

Those solutions represent the first example of nonlinear travelling wave solutions for $n<0.5$. Remarkably, the emergence of the travelling wave reduces the drag experienced by the pipe compared to the unidirectional laminar flow with the same flow rate. We demonstrated using an energy balance argument that this outcome, which is not possible for Newtonian fluids, is theoretically possible in GNF. The presence of a subtle structure near the pipe centre influences this phenomenon, indicating that accurate nonlinear analysis of shear-thinning fluid necessitates high resolution throughout the flow field.

Our original motivation was to explain the transition accompanied by an asymmetric mean flow, as mentioned in \S 1. However, our results unfortunately appear to be unrelated to this phenomenon, for the following reasons. First, in the experiments by \cite{escudier2005observations} using 0.2\% PAA, instability, characterised by an asymmetric mean flow, was observed around $Re_b=O(10^4)$.
However, our model did not show any instability under the same configuration as their experimental setup.
Second, the mean flow of our nonlinear solutions does not display the $m=1$ asymmetry seen in experiments (the axially averaged flow of the spiral solution is axisymmetric ($m=0$), while the mirror-symmetric solution has a twofold rotational symmetry ($m=2$)). Moreover, the subcricitical bifurcation of our solutions contrasts with the 
the observations of \citep{picaut2017experimental,wen2017experimental}, who reported a supercritical bifurcation.
Our review of the literature suggests that the instability and finite-amplitude travelling waves found in our study have not been observed experimentally.

\textcolor{black}{As remarked in \S 1, most real-world non-Newtonian fluids exhibit viscoelasticity and behave in more complex fashion than for the GNF model we used. 
The results of this paper strongly suggest that to fully resolve the discussion regarding the origin of the experimentally observed asymmetric mean flow (\cite{escudier2005observations,wen2017experimental,charles2024asymmetry}), reliable fully nonlinear numerical solvers that incorporate shear-thinning and viscoelastic effects, such as the White-Metzner model, are essential.}
\textcolor{black}{Note also that our results are highly limited in their applicability to blood flow. Modern models recognise blood as a thixotropic, weakly elastic viscoelastic liquid \citep{Beris21}. Furthermore, blood flow is typically neither steady nor fully developed, and the vessel walls are not rigid.}

\textcolor{black}{
Finally, we discuss the potential extension of our pipe flow stability results to other flow geometries. 
The shape of the flow cross-section is important for stability. 
For example, pipe flow and channel flow share the common characteristic of having parabolic laminar solutions in Newtonian fluids. However, pipe flow does not exhibit TS wave instabilities, and the Squire theorem does not hold. It is noteworthy that \cite{wilson1999instability} and \cite{wilson2015linear} analysed channel flow using the White-Metzner model and found new instability modes. Whether this instability exists in the inelastic limit and exhibits characteristics similar to the modes we found is an open question. \cite{nouar2007linear} investigated the stability of channel flow using the Carreau model for $n$ in the range of 0.3 to 0.7, but did not report any new instabilities. However, since they primarily focused on TS waves, it is possible that they overlooked the potential for new instability modes.} 

\textcolor{black}{
Stability analysis of shear-thinning fluid flow in a rectangular duct (see \cite{barmak2024}, for example) could be an interesting research direction. For Newtonian fluids, a TS wave appears when the aspect ratio is large, whereas when the aspect ratio is close to unity, the base flow remains linearly stable, similar to pipe flow (\cite{tatsumi1990}). In the latter case, when the Carreau model is used with sufficiently small $n$, an instability similar to those we identified may emerge. By increasing the aspect ratio, one could observe how these modes evolve and whether they remain distinguishable from TS waves.}

\backsection[Acknowledgements]{
This research was supported by the Australian Research Council Discovery Projects DP220103439 and DP230102188. \textcolor{black}{We wish to thank the referees for their constructive comments, and in particular one referee who performed an independent stability analysis.}
}

\backsection[Declaration of Interests]{
The authors report no conflict of interest.
}

\appendix
\section{Base flow computation}

We first note that the constant $q$ depends on the quadruplet $(\mu_\infty,a,n,\lambda)$, but not on $Re$. 
Integration of (\ref{base}) yields
\begin{eqnarray}
\overline{w}'F(\overline{w}')=-qr\label{baseint}
\end{eqnarray}
where $2\overline{\mu}\,=F(\overline{w}')$.
We tentatively fix $q$ at a chosen value. Then, at each designated collocation point, the value of $\overline{w}'$ can be obtained by numerically solving the implicit equation (\ref{baseint}) via the bisection method. Chebyshev integration with the imposed condition $\overline{w}(1)=0$ can be used to determine $\overline{w}(r)$. However, since $\overline{w}(0)$ is in general not equal to unity, the value of $q$ must be adjusted; this can also be done using the bisection method.
Once the correct value of $q$ is determined for given $(\mu_\infty,a,n,\lambda)$, the corresponding profiles of $\overline{w}$ and $\overline{w}'$ are also obtained. Higher order derivatives can be easily computed by
\begin{eqnarray}
\overline{w}''=-\frac{q}{F(\overline{w}')+\overline{w}'F'(\overline{w}')},\qquad
\overline{w}'''
=-(\overline{w}'')^2\frac{2F'(\overline{w}')+\overline{w}'F''(\overline{w}')}{F(\overline{w}')+\overline{w}'F'(\overline{w}')}.\label{eq:W2}
\end{eqnarray}

When $\lambda$ is large, $\overline{\mu}$ can be approximated by the power-law $(\lambda |\overline{w}'| )^{n-1}$, except for a small region around the centreline of the pipe where $\overline{w}'$ is $O(\lambda^{-1})$. 
Using the power-law form of $\overline{\mu}$, 
the solution of (\ref{base}) satisfying the no-slip boundary condition can be readily found as $\overline{w} = \{nq(2\lambda^{n-1})^{-1/n}/(n+1)\}[1-r^{1+1/n}]+\cdots$. Here, the coefficient in the curly bracket must be unity owing to our choice of velocity scale.
The aforementioned centreline region therefore exists when $r=O(\lambda^{-n})$, within which \textcolor{black}{the expansions 
\begin{eqnarray}\label{baseexp}
\overline{w}=1+\lambda^{-n-1}\overline{w}_1(\xi)+\cdots ,\qquad \overline{\mu}=\overline{\mu}_1(\xi)+\cdots, \qquad \xi=\lambda^n r,
\end{eqnarray}
hold. 
The functions $\overline{w}_1$ and $\overline{\mu}_1$ satisfy}
\begin{eqnarray}
\xi^{-1}(\xi \overline{\mu}_1 \overline{w}_1')'=-2[(n+1)/n]^n,\qquad \overline{\mu}_1(\xi)=(1+|\overline{w}_1'|^a)^{(n-1)/a},
\end{eqnarray}
and $\overline{w}_1\rightarrow -\xi^{1+1/n}$ as $\xi\rightarrow \infty$.

\section{Inviscid stability analysis}

\citet{batchelor1962analysis} showed that for an axisymmetric base flow $\overline{w}(r)$, inviscid instability is ruled out  if there exists $\alpha \in \mathbb{R}$ such that $(\overline{w}-\alpha)Q'$ does not change sign within the region of interest. Here,
\begin{eqnarray}
Q(r)=\frac{r\overline{w}'}{m^2+k^2r^2}.
\end{eqnarray}

We begin by examining the power-law velocity profile $\overline{w}=1-r^s$, where $s=1+1/n$.
It is easy to see that 
\begin{eqnarray}
Q'=-\frac{sr^{s-1}(k^2r^2(s-2)+m^2 s)}{(m^2+k^2r^2)^2}
\end{eqnarray}
vanishes at $r=0$ and $r_c$, where
\begin{eqnarray}
r_c=\frac{m}{k} \left(\frac{s}{2-s}\right)^{1/2}=\frac{m}{k} \left( \frac{n+1}{n-1} \right)^{1/2}.
\end{eqnarray}
For shear-thinning fluids ($n < 1$), $r_c$ is imaginary, implying that $Q'$ does not change sign in $r\in [0,1]$. Therefore, the stability condition can be satisfied by choosing a sufficiently large $\alpha$.
For shear-thickening fluids ($n > 1$), $r_c$ is real. However, choosing $\alpha=\overline{w}(r_c)$ allows us to show that
\begin{eqnarray}
(\overline{w}-\alpha)Q'=\frac{s(2-s)k^2r^{s-1}(r_c^2-r^2)(r_c^s-r^s)}{(m^2+k^2r^2)^2}
\end{eqnarray}
does not change sign. Therefore, for all $n$, inviscid instability is impossible for power-law fluids.

The power-law approximation is actually not needed to demonstrate the absence of inviscid instability for shear-thinning Carreau-Yasuda pipe flow.
Using (\ref{baseint}) and (\ref{eq:W2}) we can show the identity
\begin{eqnarray}
(m^2+k^2r^2)^2Q'=-\frac{q(m^2+k^2r^2)r}{F(\overline{w}')+\overline{w}'F'(\overline{w}')}+(m^2-k^2r^2)\overline{w}'\nonumber \\
=\frac{(m^2+k^2r^2)\overline{w}'F(\overline{w}')+(m^2-k^2r^2)\overline{w}'(F(\overline{w}')+\overline{w}'F'(\overline{w}'))}{F(\overline{w}')+\overline{w}'F'(\overline{w}')}.\label{mkrQ}
\end{eqnarray}
The denominator of this equation is strictly positive because
\begin{eqnarray}
F(\overline{w}')+\overline{w}'F'(\overline{w}')=2(\overline{\mu}+\check{\mu})\qquad \qquad \qquad \qquad \qquad \qquad  \nonumber \\
=\mu_\infty+(1-\mu_\infty)\{1+|\lambda \overline{w}'|^a\}^{(n-1-a)/a}\{1+n|\lambda \overline{w}'|^a\}>0.
\end{eqnarray}
Note that (\ref{eq.baseu}), (\ref{mucheck}), and $F(\overline{w}')=2\overline{\mu}$ implies that $\overline{w}'F'(\overline{w}')=2\check{\mu}$, and that $\mu_\infty<1$ by definition.
The numerator of (\ref{mkrQ}) becomes
\begin{eqnarray}
2m^2\overline{w}'F(\overline{w}')
+(m^2-k^2r^2)(\overline{w}')^2F'(\overline{w}')\qquad \nonumber \\
=
\overline{w}'
\{
2m^2(2\overline{\mu}+ \check{\mu})
-2k^2r^2 \check{\mu}
\}.
\end{eqnarray}
From (\ref{baseint}) $\overline{w}'<0$ for $r\in (0,1]$. Furthermore, for shear-thinning fluids $\check{\mu}<0$.
Therefore, if $(2\overline{\mu}+ \check{\mu})$ is positive definite, then $Q'$ does not change sign, completing the proof. This final condition can be directly shown as follows.
\begin{eqnarray}
2\overline{\mu}+ \check{\mu}=
2\mu_\infty+(1-\mu_\infty)\{1+|\lambda \overline{w}'|^a\}^{(n-1-a)/a}\{2+(n+1)|\lambda \overline{w}'|^a\}>0.
\end{eqnarray}

\section{Derivation of (\ref{energyCf})}

Here, we study the friction factor of a statistically steady flow field, such as travelling-wave solutions, at a given $Re_b$. For travelling-wave solutions, the angle brackets denote averaging over the $\theta$–$z$ directions, as in the main text; in the general case, a time average is also included.

Taking the dot product of $\mathbf{u}$ with equation (\ref{eq.non1}) and performing a spatio-temporal average yields the energy balance relation
\begin{eqnarray}
2\int^1_0\langle\mu  D\negthinspace:\negthinspace D\rangle rdr=q\int^1_0\langle w\rangle rdr.\label{muDD0}
\end{eqnarray}
We now replace the velocity scale with the bulk velocity by writing $[\mathcal{U},\mathcal{V},\mathcal{W}]=U_b^{-1}[u,v,w]$, where $U_b=2\int^1_0\langle w\rangle rdr$.
The left hand side of (\ref{muDD0}) becomes
$2U_b^2\int^1_0\langle\mu  \mathcal{D}\negthinspace:\negthinspace \mathcal{D}\rangle rdr$,
where $\mathcal{D}=D/U_b$ is the strain rate tensor expressed in terms of the rescaled velocity.

The rescaled field is then decomposed as 
\begin{eqnarray}
\mathcal{W}=\overline{\mathcal{W}}+\tilde{\mathcal{W}}_m+\tilde{\mathcal{W}}_f. \label{decomp}
\end{eqnarray}
The first two components depend only on $r$, while the last represents the fluctuation part, which is assumed to satisfy $\langle \tilde{\mathcal{W}}_f\rangle=0$. 
The first component of the mean part is defined as $\overline{\mathcal{W}}=\overline{w}/\overline{U}_b$, where $\overline{U}_b=2\int^1_0 \overline{w} rdr$, and $\overline{w}$ satisfies 
\begin{eqnarray}
(r\overline{\mu}\,\overline{w}')'=-\overline{q}.\label{baseQQ}
\end{eqnarray}
with some $\overline{q}$.
It is easy to check that 
\begin{eqnarray}
\int^1_0\tilde{\mathcal{W}}_mrdr=0,\qquad
\int^1_0\overline{\mu}\,(\overline{w}')^2rdr=\overline{q}\int^1_0\overline{w}rdr.\label{AAA}
\end{eqnarray}

Applying the decomposition to (\ref{muDD0}) yields
\begin{eqnarray}
2\int^1_0\langle\overline{\mu}  \overline{\mathcal{D}}\negthinspace:\negthinspace \overline{\mathcal{D}}\rangle rdr
+4\int^1_0\langle \bar{\mu}  \overline{\mathcal{D}}\negthinspace:\negthinspace\tilde{\mathcal{D}}_m\rangle rdr+
\beta \nonumber \\
+2\int^1_0\langle(\mu-\overline{\mu})  \mathcal{D}\negthinspace:\negthinspace\mathcal{D}\rangle rdr
=\frac{2q}{U_b},\label{BBB}
\end{eqnarray}
where
\begin{eqnarray}
\beta=2\int^1_0\langle\overline{\mu}  \tilde{\mathcal{D}}_m\negthinspace:\negthinspace\tilde{\mathcal{D}}_m\rangle rdr
+2\int^1_0\langle\overline{\mu}  \tilde{\mathcal{D}}_f\negthinspace:\negthinspace\tilde{\mathcal{D}}_f\rangle rdr.
\end{eqnarray}

The following identities can be found from (\ref{AAA}).
\begin{eqnarray*}
2\int^1_0\langle \bar{\mu}  \overline{\mathcal{D}}\negthinspace:\negthinspace\tilde{\mathcal{D}}_m\rangle rdr=\int^1_0 \bar{\mu}  \overline{\mathcal{W}}'\tilde{\mathcal{W}}_m' rdr=-\int^1_0 (r\bar{\mu}  \overline{\mathcal{W}}')'\tilde{\mathcal{W}}_m dr=\frac{\overline{q}}{\overline{U}_b}\int^1_0 \tilde{\mathcal{W}}_m rdr=0,\\
2\int^1_0\langle\overline{\mu}  \overline{\mathcal{D}}\negthinspace:\negthinspace\overline{\mathcal{D}}\rangle rdr=\int^1_0\overline{\mu}  \overline{\mathcal{W}}' \overline{\mathcal{W}}' rdr
=-\int^1_0(r\overline{\mu}  \overline{\mathcal{W}}')' \overline{\mathcal{W}}' dr=\frac{\overline{q}}{\overline{U}_b}\int^1_0 \overline{\mathcal{W}} rdr=\frac{2\overline{q}}{\overline{U}_b}.
\end{eqnarray*}
Substituting them to (\ref{BBB}), we get:
\begin{eqnarray}
0\leq \beta=-2\int^1_0\langle(\mu-\overline{\mu})  \mathcal{D}\negthinspace:\negthinspace\mathcal{D}\rangle rdr+\frac{2q}{U_b}-\frac{2\overline{q}}{\overline{U}_b}.\label{C6}
\end{eqnarray}
On the right hand side, $q$ and $\overline{q}$ can be rewritten using
\begin{eqnarray}
C_f=\frac{4q}{Re U_b^2}, \qquad \overline{C}_f=\frac{4\overline{q}}{\overline{Re}\,\overline{U}_b^2}.\label{C7}
\end{eqnarray}
Here $\overline{Re}$ is adjusted to satisfy
\begin{eqnarray}
Re_b=\frac{4\overline{Re}}{\overline{\mu} |_{r=1}}\int^1_0 \langle \overline{w} \rangle rdr=\frac{4Re}{\langle \mu \rangle |_{r=1}}\int^1_0 \langle w \rangle rdr.\label{C8}
\end{eqnarray}
Equation (\ref{energyCf}) can be obtained by combining equations (\ref{C6}), (\ref{C7}), and (\ref{C8}).

\section{FIK identity for GNF}
The FIK identity can be easily obtained by multiplying the mean $z$-momentum equation by the base flow and integrating over the domain.
\begin{eqnarray}
\int^1_0\overline{w}\partial_r(r\langle uw\rangle ) dr=\frac{1}{Re}\int^1_0\overline{w}\partial_r(r\langle \mu (\partial_z u+\partial_rw)\rangle )dr+\frac{q}{Re}\int^1_0\overline{w}rdr.
\end{eqnarray}
By performing integration by parts and applying the rescaling introduced in the previous section,
\begin{eqnarray}
-\int^1_0\overline{\mathcal{W}}'\langle \mathcal{U}\mathcal{W}\rangle  rdr=-\frac{1}{ReU_b}\int^1_0\overline{\mathcal{W}}'r\langle \mu (\partial_z\mathcal{U}+\partial_r\mathcal{W})\rangle dr+\frac{q}{2ReU_b^2}.
\end{eqnarray}
Further application of the  decomposition (\ref{decomp}) yields
\begin{eqnarray}
-\int^1_0\overline{\mathcal{W}}'\langle \tilde{\mathcal{U}}_f\tilde{\mathcal{W}}_f\rangle  rdr=
-\frac{1}{ReU_b}\int^1_0\overline{\mathcal{W}}'r\langle (\mu-\overline{\mu}) (\partial_z\mathcal{U}+\partial_r\mathcal{W})\rangle dr \nonumber \\
-\frac{1}{ReU_b}\frac{\overline{q}}{2\overline{U}_b}
+\frac{q}{2ReU_b^2}.\label{FIK}
\end{eqnarray}
Here we have used
\begin{eqnarray*}
\int^1_0\overline{\mathcal{W}}'r \overline{\mu} (\overline{\mathcal{W}}'+\tilde{\mathcal{W}}_m') dr=-\int^1_0(\overline{\mathcal{W}}'r \overline{\mu})' (\overline{\mathcal{W}}+\tilde{\mathcal{W}}_m) dr
=\frac{\overline{q}}{\overline{U}_b}\int^1_0 \overline{\mathcal{W}} rdr
=\frac{\overline{q}}{2\overline{U}_b}.
\end{eqnarray*}
Upon using (\ref{C7}), and (\ref{C8}), the last two terms of (\ref{FIK}) can be written in terms of the friction factors. Finally, we get
\begin{eqnarray}
\frac{8}{ReU_b}\int^1_0\overline{\mathcal{W}}'\langle (\mu-\overline{\mu}) (\partial_z\mathcal{U}+\partial_r\mathcal{W})\rangle rdr-8\int^1_0\overline{\mathcal{W}}'\langle \tilde{\mathcal{U}}_f\tilde{\mathcal{W}}_f\rangle  rdr\nonumber \\
=C_f-\frac{\overline{\mu} |_{r=1}}{\langle \mu \rangle|_{r=1}}\overline{C}_f=\overline{C}_f\left (\left . \left (\frac{\langle \mu \rangle-\overline{\mu} }{\langle \mu \rangle}\right)\right |_{r=1}+\Delta C_f \right ).\label{FIK2}
\end{eqnarray}
The first term on the right hand side vanishes for a Newtonian fluid, and the standard FIK identity is recovered.
\bibliographystyle{jfm}  
\bibliography{Reference}  

\end{document}